\documentclass[useAMS,usenatbib,twocolumn]{mnras}
\usepackage{natbib}
\usepackage{graphicx}
\usepackage{times}
\usepackage{rotate}
\usepackage{hyperref}
\usepackage{url}

\newcommand{\Ha}{$\rm H\alpha$}

\newcommand{\ds}{\ensuremath{\Delta\Sigma}}

\newcommand{\merianmass}{$8<\rm \log M_*/M_\odot<9$}
\newcommand{\iha}{$i-\rm N708$}
\newcommand{\roiii}{$r-\rm N540$}

\usepackage[multiple]{footmisc}

\begin{document}

\title[The Merian Filter Set]{The Merian Survey: Design, Construction, and Characterization of a Filter Set Optimized to Find Dwarf Galaxies and Measure their Dark Matter Halo Properties with Weak Lensing}

 \author[Y. Luo et al.]{Yifei Luo$^{1}$\thanks{E-mail: yifeiluo@ucsc.edu}, Alexie Leauthaud$^{1}$, Jenny Greene$^{2}$, Song Huang$^{1,2,3}$, Erin Kado-Fong$^{2,4}$, 
 \newauthor
 Shany Danieli$^{2}$, Ting S. Li$^{5}$, Jiaxuan Li$^{2}$, Diana Blanco$^{1}$, Erik J. Wasleske$^{6}$, Joseph Wick$^{1}$,  
 \newauthor
 Abby Mintz$^{2}$, Runquan Guan$^{1}$, Annika H. G. Peter$^{7,8,9,10}$, Vivienne Baldassare$^{6}$, 
 \newauthor
 Alyson Brooks$^{11,12}$,  Arka Banerjee$^{13}$, Joy Bhattacharyya$^{7,9}$, Zheng Cai$^{3}$, Xinjun Chen$^{1}$,    
 \newauthor
Jim Gunn$^{2}$, Sean D. Johnson$^{14}$, Lee S. Kelvin$^{2}$, Mingyu Li$^{3}$, Xiaojing Lin$^{3}$, Robert Lupton$^{2}$, 
 \newauthor 
Charlie Mace$^{7,8,9}$, Gustavo E. Medina$^{5}$, Justin Read$^{15}$, Rodrigo Córdova Rosado$^{2}$, 
 \newauthor 
Allen Seifert$^{1}$
\\
$^1$Department of Astronomy and Astrophysics, University of California, Santa Cruz, 1156 High Street, Santa Cruz, CA 95064 USA\\
$^2$Department of Astrophysical Sciences, Princeton University, 4 Ivy Lane, Princeton, NJ 08544, USA\\
$^3$Department of Astronomy, Tsinghua University, Beijing 100084, People’s Republic of China\\
$^4$Physics Department, Yale Center for Astronomy \& Astrophysics, PO Box 208120, New Haven, CT 06520, USA\\
$^5$Department of Astronomy and Astrophysics, University of Toronto, 50 St. George Street, Toronto ON, M5S 3H4, Canada\\
$^6$Department of Physics and Astronomy, Washington State University, Pullman, WA 99163, USA\\
$^7$The Ohio State University, Department of Physics, Columbus, OH 43210, USA\\
$^8$The Ohio State University, Department of Astronomy, Columbus, OH 43210, USA\\
$^{9}$Center for Cosmology and Astroparticle Physics, 191 West Woodruff Avenue, Columbus, OH 43210, USA\\
$^{10}$School of Natural Sciences, Institute for Advanced Study, Princeton, NJ 08540, USA\\
$^{11}$Department of Physics and Astronomy, Rutgers, The State University of New Jersey, 136 Frelinghuysen Road, Piscataway, NJ 08854, USA\\
$^{12}$Center for Computational Astrophysics, Flatiron Institute, 162 Fifth Avenue, New York, NY 10010, USA\\
$^{13}$Department of Physics, Indian Institute of Science Education and Research,
Homi Bhabha Road, Pashan, Pune 411008, India\\
$^{14}$Department of Astronomy, University of Michigan, Ann Arbor, MI 48109, USA\\
$^{15}$Department of Physics, University of Surrey, Guildford, GU2 7XH, United Kingdom\\
}
\maketitle

\label{firstpage}

\begin{abstract} 
The Merian survey is mapping $\sim$ 850 degrees$^2$ of the Hyper Suprime-Cam Strategic Survey Program (HSC-SSP) wide layer with two medium-band filters on the 4-meter Victor M. Blanco telescope at the Cerro Tololo Inter-American Observatory, with the goal of carrying the first high signal-to-noise (S/N) measurements of weak gravitational lensing around dwarf galaxies. This paper presents the design of the Merian filter set: N708 ($\lambda_c = 7080$\AA, $\Delta\lambda = 275$\AA) and N540 ($\lambda_c = 5400$\AA, $\Delta\lambda = 210$\AA). The central wavelengths and filter widths of N708 and N540 were designed to detect the \Ha\ and $\rm [OIII]$ emission lines of galaxies in the mass range \merianmass\ by comparing Merian fluxes with HSC broad-band fluxes. Our filter design takes into account the weak lensing S/N and photometric redshift performance. Our simulations predict that Merian will yield a sample of $\sim$ 85,000 star-forming dwarf galaxies with a photometric redshift accuracy of $\sigma_{\Delta z/(1+z)}\sim 0.01$ and an outlier fraction of $\eta=2.8\%$ over the redshift range $0.058<z<0.10$. With 60 full nights on the Blanco/Dark Energy Camera (DECam), the Merian survey is predicted to measure the average weak lensing profile around dwarf galaxies with lensing S/N $\sim 32$ within $r<0.5$ Mpc and lensing S/N $\sim 90$ within $r<1.0$ Mpc. This unprecedented sample of star-forming dwarf galaxies will allow for studies of the interplay between dark matter and stellar feedback and their roles in the evolution of dwarf galaxies.
\end{abstract}

 
\section{Introduction}

Dwarf galaxies are unique probes of the nature of dark matter and baryonic processes such as stellar feedback \citep[e.g.][]{Collins:2022,Lelli:2022,Battaglia:2022,Sales:2022}. Extensive studies of dwarf galaxies in the Local Group and beyond have revealed a considerable scatter in many of their fundamental properties \citep[e.g.][]{Bullock:2017}. Reliably characterizing both the dark matter and baryonic components of these galaxies is key to establishing a complete picture of dark matter on small scales, understanding the threshold of galaxy formation, and studying dark matter in dwarf galaxies. 

For several decades, $\Lambda$CDM simulations have struggled to reproduce the properties of observed dwarf galaxies \citep[``Small-Scale Challenges'', ][]{Bullock:2017}. These challenges include the ``Core-Cusp'' \citep[e.g.][]{Flores:1994,Moore:1994}, ``Missing Satellites'' \citep[e.g.][]{Klypin:1999,Moore:1999}, and ``Too-Big-to-Fail'' \citep[e.g.][]{Read:2006,Boylan-Kolchin:2011} problems. In recent years, simulations have produced more realistic dwarf galaxies due to improvements in resolution and more physically-motivated treatments of stellar feedback \citep[e.g.,][]{Brooks:2014,Wetzel:2016,Brooks:2017}. However, it remains unclear whether feedback prescriptions are realistic enough. Indeed, reproducing both the distribution of star-formation rates, and the observed scatter in dwarf rotation curves, has proven challenging for $\Lambda$CDM simulations. This is known as the ``diversity'' problem \citep[e.g.,][]{Oman:2015, Santos-Santos:2018, Roper:2023}. Non-CDM theories of dark matter such as self-interacting, warm, and fuzzy dark matter \citep[e.g.][]{Fry:2015, Kamada:2017,Robles:2017,Fitts:2019,Nadler:2021} have been proposed as solutions to the diversity problem. Alternatively, it is possible that the star formation feedback sub-grid models in simulations are not yet accurate enough. Or, it could be that data themselves have been misinterpreted. We know of at least one example, IC1613, where a combination of stellar and HI gas kinematics show that its inclination is systematically biased, driving an apparent scatter in the rotation curve \citep[e.g.][]{Oman:2016,Read:2016,Collins:2022}. It is not clear, however, whether such effects can explain all the observed diversity.

Improved measurements of the stellar-to-halo mass relation (SHMR) for dwarf galaxies are important not only to understand galaxy formation and evolution, but also to disentangle baryonic effects from non-CDM theories of dark matter \citep[e.g.][]{Nadler:2020, Munshi:2021, Danieli:2022}. The SHMR is poorly constrained on dwarf galaxy scales, even for massive dwarf galaxies (e.g., \merianmass), which are easier to detect compared to fainter (lower mass) dwarfs \citep[e.g.][]{Miller:2014,Hudson:2015, Read:2017, Posti:2019, Read:2019b}. The mass range \merianmass\ is a ``sweet spot'' for detecting correlations between the properties of a dwarf's baryonic matter, the slope of its inner dark matter halo, and its overall dark matter halo mass \citep[e.g.,][]{DiCintio:2014a,DiCintio:2014b,Relatores:2019}. Zoom-in hydrodynamical simulations, such as NIHAO and FIRE-2, have revealed that baryonic feedback is able to efficiently produce core-like density profiles for galaxies in the mass range \merianmass\ \citep[e.g.][]{Mashchenko:2008,Governato:2010,Governato:2012,Pontzen:2012,Teyssier:2013,Wetzel:2016,Tollet:2016}. We now have compelling observational evidence for such cusp-core transformations occurring in nearby dwarf galaxies \citep{Read:2019a, Bouche:2022, DeLeo:2023}. Finally, $\Lambda$CDM simulations with pure dark matter and observations also show differences in terms of the inner dark matter halo kinematics at $\rm M_*<10^{9} M_{\odot}$ \citep{Klypin:2015,Papastergis:2015}. Constraining the SHMR at these mass scales is therefore of great interest.

Measurements of the dark matter halos of dwarf galaxies are key to understanding the dwarf population. Rotation curves have been used to constrain dark matter halos for decades \citep[e.g.,][]{Rubin:1978,Rubin:1980,Boylan-Kolchin:2012,Oh:2015}. However, rotation curves only constrain the inner regions of dark matter halos, where most of the baryons are located (e.g., stars, gas, globular clusters). A typical $\rm M_*=10^{8.5}M_{\odot}$ dwarf galaxy has a light-weighted effective radius of $\sim$ 1 kpc, if not smaller \citep{Eigenthaler:2018}. The halo radius R$_{\rm 200m}$\footnote{The radius enclosing an overdensity of 200 times the mean matter density of the universe.} for a $\rm M_*=10^{8.5}M_{\odot}$ galaxy is $90-150$ kpc in simulations \citep[e.g.,][]{Kravtsov:2013}. This means that the radial scales probed by rotation curves are typically a factor of $\sim$20-30 smaller than the R$_{\rm 200m}$. Any halo mass estimate from rotation curves is therefore an extrapolation that relies on assumptions about the shape of the dark matter profile \citep[e.g.,][]{Buckley:2018,McQuinn:2022}. For this reason, methods for measuring the full dark matter halos of dwarfs will be of great value.

A powerful way to directly probe total halo masses is via weak gravitational lensing. In particular, the ``galaxy-galaxy lensing'' technique measures the average weak lensing signal from background ``source'' galaxies around a sample of foreground ``lens'' galaxies \citep[e.g.,][]{Brainerd:1996}. Existing weak lensing measurements have been limited to galaxies with $\rm M_*>10^{9}M_{\odot}$ \citep[e.g.,][]{Leauthaud:2012,Hudson:2015,Dvornik:2022}.  \citet{Leauthaud:2020} recently predicted that lensing surveys such as the HSC Strategic Survey Program \citep[HSC-SSP, e.g.][]{Aihara:2018}, $\rm Euclid$ \citep{Laureijs:2011}, the $\rm Roman$ Space Telescope \citep{Spergel:2013}, and the Legacy Survey of Space and Time \citep[LSST,][]{Ivezi:2019} with the Rubin Observatory, will be sufficiently deep and wide to provide large samples of dwarf galaxies out to $z\sim0.2$, enabling measurements of galaxy-galaxy lensing for dwarf galaxies with high significance. 

This paper focuses on a weak lensing survey for dwarf galaxies using the existing HSC-SSP survey. The HSC-SSP wide layer is mapping $\sim$1000 deg$^2$ down to i $\sim$ 26~mag in $grizy$ \citep[][]{Aihara:2018}. The $5\sigma$ depths for $g$, $r$, $i$, $z$, and $y$ in the HSC-wide field are 26.5, 26.1, 25.9, 25.1, 24.4 mag, respectively \citep[][]{Aihara:2018}. The HSC-SSP survey has both the required depth and area coverage to select large samples of low redshift dwarfs (see more details in Section~\ref{data_COSMOS}). Note that several other existing surveys on the Dark Energy Camera (DECam) also have the ability to detect large samples of dwarf galaxies for lensing. For example, the Dark Energy Survey \citep[DES,][]{Abbott:2021} covers $\sim$5000 deg$^2$ down to $i\sim23.8$ ($10\sigma$ depth). The DECam Local Volume Exploration survey \citep[DELVE,][]{Drlica-Wagner:2022} covers $\sim$17,000 deg$^2$ down to $i\sim23.5$ ($5\sigma$ depth). Both of these two surveys could also form large samples of dwarf galaxies below $z\sim0.2$ that are comparable to HSC-SSP or even larger, benefiting from their wide survey areas. However, our decision to utilize the HSC-SSP survey in this work is due to several other reasons. First, the weak lensing measurements require not only a large sample of lens galaxies, but also a large sample of source galaxies. Accurate galaxy shape measurements of these distant sources are crucial for lensing analysis. Imaging surveys with a better Point Spread Function (PSF) have less smearing effect, which could provide more precise shape measurements. Among the three existing deep and wide surveys mentioned above, HSC-SSP has the best PSF (i-band median seeing $\sim$0.61 arcsec) compared to DES (i-band median seeing $\sim$0.88 arcsec) and DELVE (i-band median seeing $\sim$1.02 arcsec). Furthermore, HSC-SSP is also deeper than the other two surveys, which means it can detect more source galaxies at higher redshifts. The high number density of the source galaxies in HSC-SSP will also help with the lensing measurements. Therefore, we choose to use the HSC-SSP survey to compile our dwarf galaxy sample for lensing in this work.

Although HSC-SSP is deep enough to detect a large sample of dwarf galaxies, their broad-band photometric redshifts are not optimized for dwarf galaxies at low redshifts. Weak lensing measurements usually need redshifts for lens galaxies with an accuracy of $\Delta z<0.2$, but the typical photo-$z$ precision at $z\sim 0.1$ from broad-band surveys with $grizy$ bands (such as HSC) is $\Delta z\sim0.5$ \citep{Tanaka:2018,Speagle:2019}. This is because the HSC filter set does not span the 4000\AA\ break at low redshifts. Furthermore, most dwarf galaxies in the mass range \merianmass\ are star-forming \citep{Geha:2012,Mao:2021} and do not have a prominent 4000\AA\ break. Thus, the inclusion of $u$-band data would not guarantee accurate redshifts for star-forming dwarf galaxies. In some cases, imaging-based machine learning methods can be used to improve redshift estimates. For example, \citet{Darragh-Ford:2022} trained a Convolutional Neural Network (CNN) with broad-band images from the Dark Energy Camera Legacy Survey \citep[DECaLS, ][]{Dey:2019} to select dwarf galaxies at $z<0.03$. This method achieved a 90\% completeness, but with only 20\% efficiency. Although their efficiency is higher than traditional photometric methods ($\sim$ 1\%), this method still requires spectroscopic follow-up observations.

Narrow and medium band filters have also been used with great success to improve redshift estimates.  This approach takes advantage of the fact that narrow/medium-band filters can be used to isolate specific spectral features, such as emission lines or absorption features, which provide additional redshift information. By integrating these features into the photometric redshift estimation procedure, redshift estimates can be significantly improved, particularly for galaxies with strong emission lines. This technique has been successfully applied in a number of imaging surveys, such as the COSMOS survey \citep{Ilbert:2009,Laigle:2016}, the ALHAMBRA survey \citep{Moles:2008}, the PAU survey \citep{Benitez:2009}, the J-PAS survey \citep{Benitez:2014}, the S-PLUS survey \citep{deOliveira:2019}, the SILVERRUSH survey \citep{Ouchi:2018}, the LAGER survey \citep{Zheng:2017,Zheng:2019}, and the ODIN survey \citep{Lee:2024}. In this paper, we present the design and optimization of a dual filter system on DECam to detect the emission lines of dwarf galaxies in the range \merianmass ~in the HSC-SSP fields. 

The Merian survey\footnote{Honoring 17th century Maria Sibylla Merian (1647, 1717), one of the first female entomologists and naturalists. Her fascination with the world of tiny things combined with unique observational skills led to a number of important discoveries, including the previously unknown metamorphosis of caterpillars into butterflies.}\footnote{\url{https://merian.sites.ucsc.edu/}} is an ongoing large imaging survey with 60 nights on the Victor M. Blanco 4-m telescope at the Cerro Tololo Inter-American Observatory (CTIO). Merian uses imaging in two medium-band filters to build a sample of dwarf galaxies at $0.058<z<0.10$. Merian will image $\sim$ 850 deg$^2$ of the HSC-SSP survey and is expected to detect $\sim$ 85,000 dwarf galaxies in the mass range \merianmass. This paper presents the design of the Merian filters. Merian will yield the first high S/N weak lensing profile measurements for dwarf galaxies. HSC-SSP imaging data will also provide exquisite measurements of dwarf sizes, shapes, stellar masses, and star-formation rates. The Merian Survey design, observing strategy, data reduction, and preliminary data products will be presented in a companion paper (Danieli et al., in prep).

This paper is organized as follows. The data and sample used to design the Merian filters are described in Section~\ref{data_COSMOS}. The filter design methodology is presented in Section~\ref{methodology}. Section~\ref{results} shows the filter optimization results and the expected photo-$z$ performance. Section~\ref{filter_construction} describes the construction of the Merian filter set, and shows the performance of two Merian filters using an early reduction of the Merian data. Section~\ref{conclusions} presents a summary and our conclusions. We adopt a flat $\Lambda$CDM cosmology with $H_0$ = 70 km $\rm s^{-1}\ Mpc^{-1}$, $\Omega_m$ = 0.3 and $\Omega_\Lambda$ = 0.7. We use physical units for the galaxy-galaxy lensing observable $\Delta\Sigma$. All magnitudes are in the AB system \citep{Oke1983}.

\section{Data Used for Filter Design}\label{data_COSMOS}

In this section, we introduce the data set used to design the Merian dual filter system. We show the basic observational properties of a mass complete sample of dwarf galaxies within a target mass and redshift range drawn from the COSMOS2015 catalog. We also build a model SED library for our target dwarf galaxies by fitting the 30-band photometry from the COSMOS2015 catalog to predict the fluxes of these dwarf galaxies in different filter designs.

\subsection{The COSMOS catalog}

To predict the S/N of fluxes of dwarf galaxies in different filter designs, we first need a sample of dwarf galaxies (our truth table) to fully understand their observational properties. This truth table should contain extensive photometric data, redshift measurements, and be deep enough to be complete in the stellar mass and redshift range of interest (\merianmass\ and $z<0.2$). 

We use the COSMOS2015 catalog as our baseline for understanding the properties of dwarfs to inform our analysis. The COSMOS2015 catalog \citep{Laigle:2016} offers a wealth of deep multi-wavelength photometric data over 2 deg$^2$ in the COSMOS field \citep{Scoville:2007} and contains nearly 500,000 objects. It combines photometry from X-ray \citep[Chandra,][]{Civano:2016}, near-UV \citep[GALEX,][]{Zamojski:2007}, optical \citep[CFHT/MegaCam, Subaru COSMOS-20, and Hyper-Suprime-Cam,][]{Taniguchi:2007,Taniguchi:2015,Miyazaki:2018}, near-IR \citep[UltraVISTA,][]{McCracken:2012}, mid-IR \citep[MIPS/Spitzer,][]{LeFloch:2009}, and far-IR \citep[PACS/Herschel and SPIRE/Herschel,][]{Lutz:2011,Oliver:2012}. The UV and optical photometry achieve a 3$\sigma$ depth of at least 25 mag. We obtain photometric redshifts ($z$PDF), stellar masses (MASS$_{\rm MED}$), half-light radii, and star-formation rates (SFR) from the COSMOS2015 catalog \citep{Laigle:2016}. We start with galaxies ({\tt TYPE=0}) within the 2 deg$^2$ COSMOS area from the COSMOS2015 catalog ({\tt FLAG\_COSMOS=1}), and then select galaxies with reliable photometric redshifts in the UltraVISTA area ({\tt FLAG\_HJMCC=0}), not influenced by artifacts around saturated objects ({\tt FLAG\_PETER=0}). We obtain the half-light radii measured from high resolution {\it HST}/ACS F814W COSMOS images  \citep{Leauthaud:2007}.

\begin{figure}
\begin{center}
\includegraphics[width=8cm]{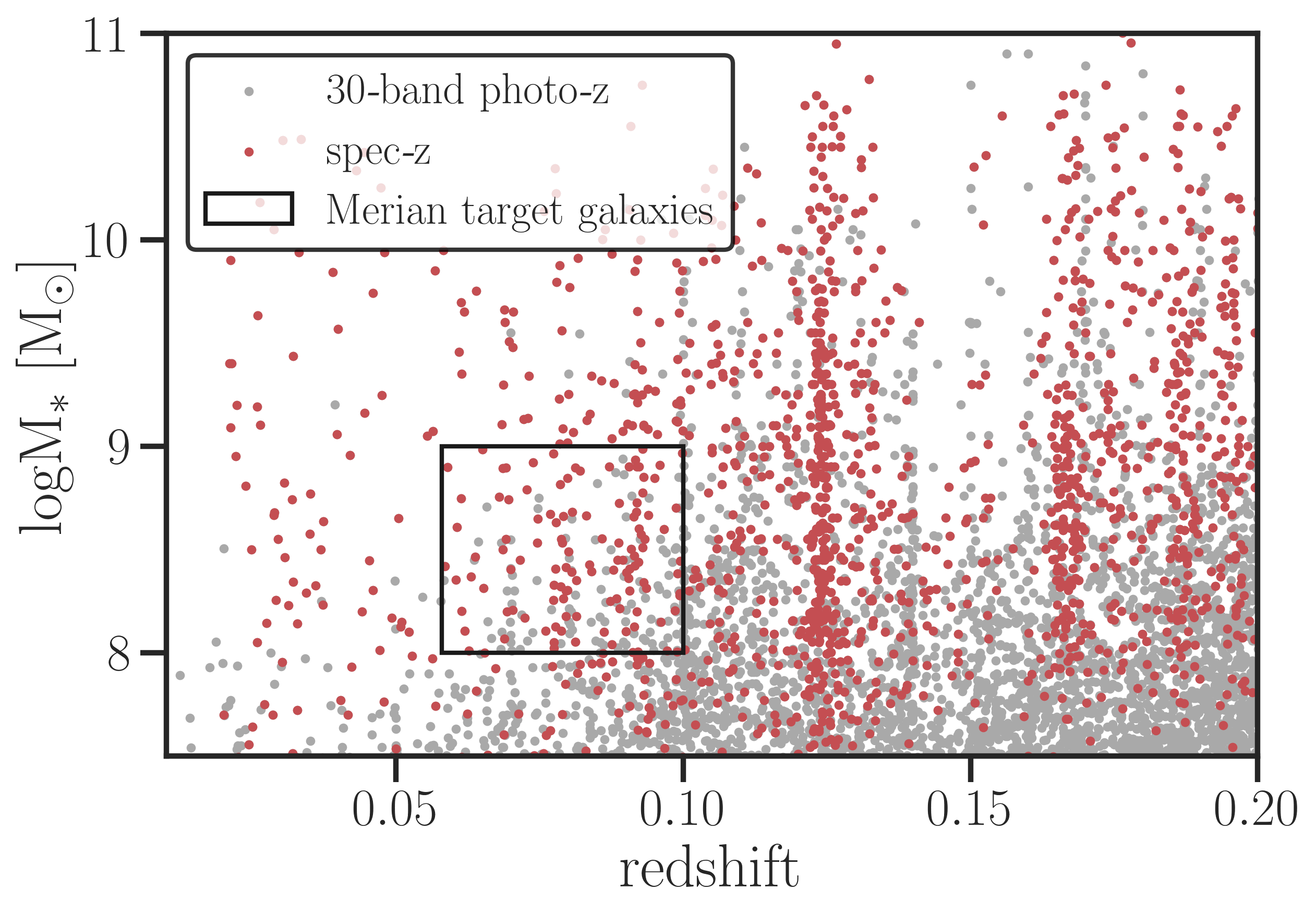}
\caption{Stellar mass vs. redshift for galaxies in the COSMOS2015 catalog. Red points indicate galaxies with spectroscopic redshifts and grey points correspond to galaxies with COSMOS 30-band photometric redshifts. The black box indicates the redshift and stellar mass range targeted by the Merian survey and corresponds to the filter design presented in this paper (see details in Section~\ref{results}).}
\label{mass_z}
\end{center}
\end{figure}

The COSMOS2015 catalog provides photometric redshift measurements with high accuracy ($\sigma_{\Delta z/(1+z)}\sim 0.01$ for both star-forming and quiescent galaxies with $i\sim 23$, and $\sigma_{\Delta z/(1+z)}\sim 0.02$ for fainter objects with $23<i<24$), along with stellar mass and star-formation rate estimates \citep{Laigle:2016}. These properties are derived with template fitting using the Spectral Energy Distribution (SED) fitting code LEPHARE \citep{Arnouts:2002, Ilbert:2006}, assuming a \citet{Chabrier:2003} Initial Mass Function (IMF), and a combination of exponentially declining and delayed star-formation histories (SFH) ($\rm SFR \sim \tau^{-2}te^{-t/\tau}$ with e-folding time $\tau$). Figure~\ref{mass_z} shows stellar mass vs. 30-band photometric redshift for galaxies from the COSMOS2015 catalog. 

\begin{figure}
\begin{center}
\includegraphics[width=8cm]{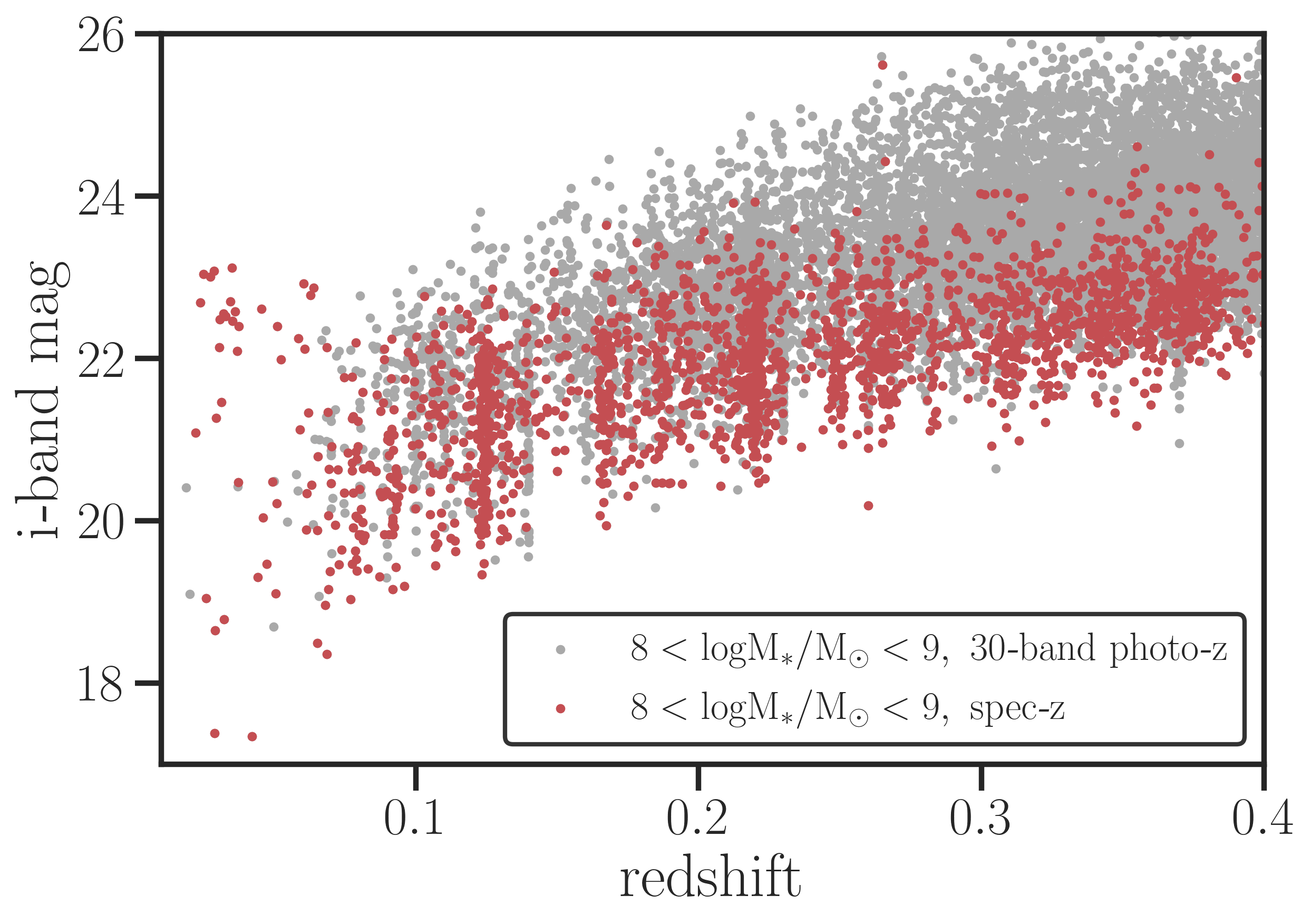}
\caption{$i$-band magnitude vs. redshift for dwarf galaxies with \merianmass\ in the COSMOS2015 catalog. Red points indicate galaxies with spectroscopic redshifts and grey points correspond to galaxies with COSMOS 30-band photometric redshifts.}
\label{imag_z}
\end{center}
\end{figure}

We also obtained spectroscopic redshifts (spec-$z$'s) from the COSMOS team (Salvato et al., in prep). This catalog includes spectroscopic redshifts from several spectroscopic surveys, including the $z$COSMOS survey \citep{Lilly:2007}, the VIMOS Ultra Deep Survey \citep[VUDS,][]{LeFevre:2015}, the Complete Calibration of the Color–Redshift Relation Survey \citep[C3R2,][]{Masters:2019}, the DEIMOS 10K Spectroscopic Survey \citep{Hasinger:2018}, and the FMOS-COSMOS Survey \citep{Kashino:2019}. Approximately 30 percent of galaxies at $z<0.2$ and $\rm \log M_*/M_\odot>8$ have spectroscopic redshift measurements (indicated by red points in Figure~\ref{mass_z}). Figure~\ref{imag_z} shows  $i$-band magnitude vs. redshift for dwarf galaxies in our target mass range \merianmass\  color-coded according to their redshift type. As can be seen, spec-$z$'s are incomplete for faint dwarf galaxies (e.g. $i>23$). The selection functions of existing spec-$z$ samples are complicated, i.e. they are not representative in colors/brightness, since most of the large spectroscopic surveys in COSMOS targeted galaxies at higher redshifts. In the remainder of this paper, when using the COSMOS2015 redshifts ($\rm z_{COSMOS}$), we will use the spectroscopic redshift when available, and the photometric redshift $z$PDF (median of the likelihood distribution) otherwise.


\begin{figure*}
\begin{center}
\includegraphics[width=15cm]{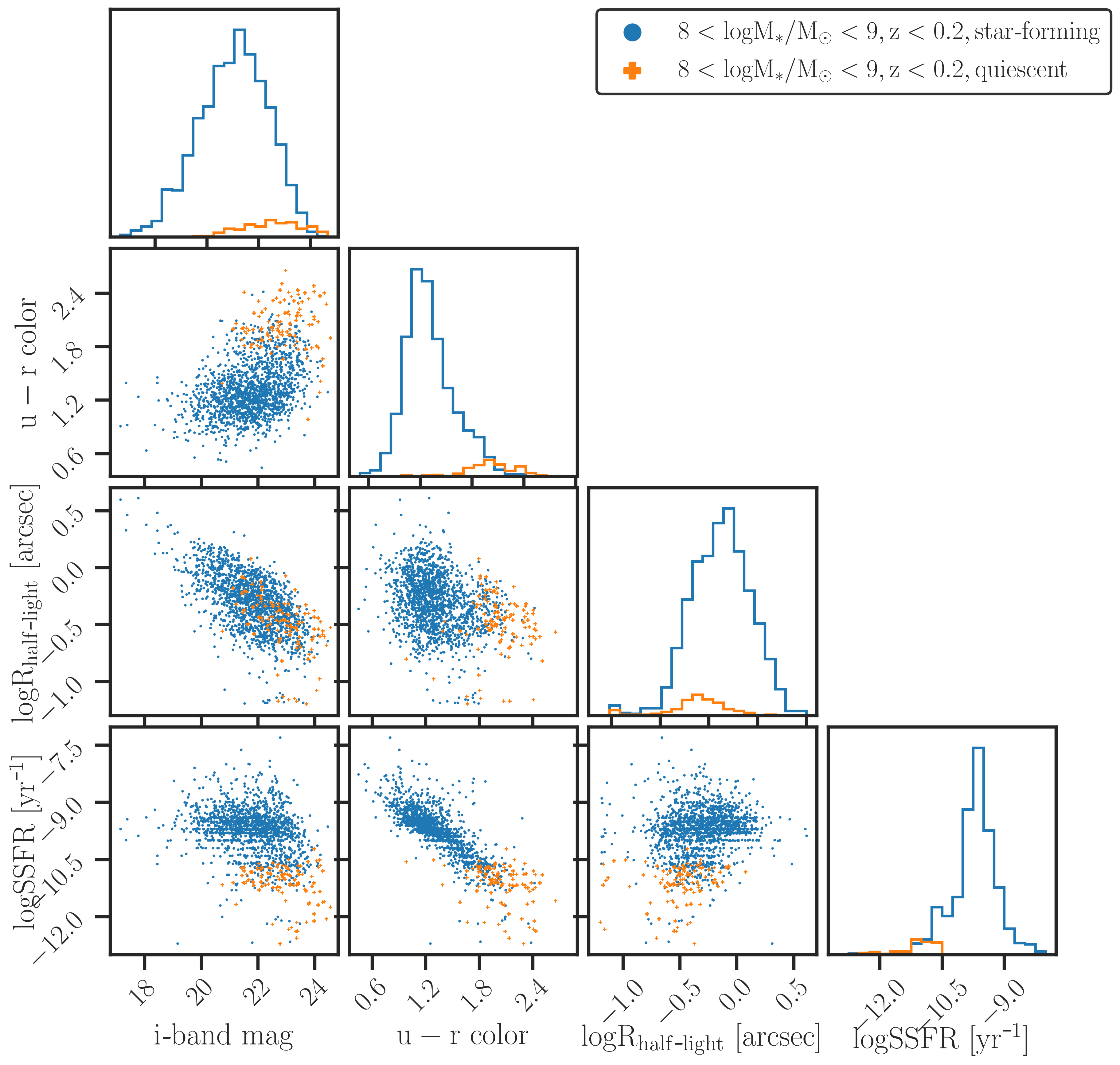}
\caption{Distributions of four basic properties ($i$-band magnitude, $u-r$ color, half-light radius, and specific star-formation rate) for dwarf galaxies with \merianmass\ and $\rm z_{COSMOS}<0.2$ (our Initial Dwarf Sample) from the COSMOS2015 catalog. Blue points (histograms) and orange crosses (histograms) correspond respectively to star-forming and quiescent dwarf galaxies in the Initial Dwarf Sample according to the classification in the COSMOS2015 catalog.}
\label{dwarf_hist}
\end{center}
\end{figure*}

\subsection{Starting Parameter Space for Filter Design}

Our goal is to work in the mass and redshift range where HSC is complete and unaffected by surface brightness limits. \citet{Leauthaud:2020} showed that detections in both COSMOS and HSC would be primarily affected by surface brightness effect for galaxies with $\rm \log M_*/M_\odot\sim 8$ at $z>0.2$ and that galaxies with $\rm \log M_*/M_\odot\sim 7.5$ are $\sim$90\% complete at $z=0.2$ \citep[see Figure 1 in][]{Leauthaud:2020}. Therefore, we focus on galaxies in the target mass and redshift range: \merianmass\ and $z<0.2$ where HSC should not be affected by surface brightness limits.

As shown in Figure~\ref{imag_z}, dwarf galaxies with \merianmass\ at $\rm z_{COSMOS}<0.2$ can reach magnitudes as faint as $i\sim 24$. It is important to note that the final Merian survey target redshift range does not extend to $z=0.2$, but rather spans $0.058<z<0.1$. Consequently, most of the Merian target dwarfs are actually brighter than $i\sim23$, as shown in Figure~\ref{imag_z} (see other properties for the final Merian target dwarfs in Figure~\ref{dwarf_hist_merian} in Section~\ref{results}). Nonetheless, for the filter design process, we started with dwarf galaxies at $z<0.2$. For this, it was important to have a sample of dwarfs with deep enough data (down to $i\sim 24$) to use as a truth table. In reality, very few large imaging surveys are complete down to $i\sim 24$, and none of the large spectroscopic surveys are deep enough to form a library of SEDs for our target dwarf galaxies. For instance, the Sloan Digital Sky Survey (SDSS) and Galaxy And Mass Assembly (GAMA) surveys are only complete down to $r<17.77$ and $r<19.8$ mag (much shallower than $i\sim 24$), respectively \citep{York:2000,Liske:2015}. Deeper spectroscopic surveys do exist, such as VVDS ($I_{\rm AB}<22.5$), PRIMUS ($i_{\rm AB}<23.5$), and DEEP2 ($R_{\rm AB}<24.1$) \citep{LeFevre2013,Coil:2011,Newman:2013}, but deeper surveys usually use specific color selections to target high-redshift galaxies, thereby excluding some dwarf galaxies in our target mass and redshift range. For this reason, we choose to use the COSMOS2015 catalog as our initial truth table and to build our dwarf SED library. We select 1,822 dwarf galaxies with \merianmass\ at $\rm z_{COSMOS}<0.2$ from the COSMOS2015 catalog, of which 639 have spectroscopic redshifts. We refer to this sample as the ``Initial Dwarf Sample" and the basic properties of this sample are presented below.

\subsection{Initial Dwarf Sample}\label{Initial_Dwarf_Sample}

Figure~\ref{dwarf_hist} shows the distributions of $i$-band apparent magnitude, $u-r$ color, half-light radius in arcseconds, and specific star-formation rate (sSFR) for star-forming and quiescent galaxies for the Initial Dwarf Sample. The star-forming/quiescent classification used here is from the COSMOS2015 catalog, and is based on absolute $\rm NUV-r$ and $\rm r-J$ colors \citep{Laigle:2016}. The $K$-correction in each band is estimated from the 30-band SED fitting. Quiescent objects have $\rm NUV-r>3(r-J)+1$ and $\rm NUV-r>3.1$. According to this classification, 95\% of dwarf galaxies in the Initial Dwarf Sample are star-forming (blue points/histograms in Figure~\ref{dwarf_hist}). Similarly, nearly 95\% of dwarf galaxies in the Initial Dwarf Sample have an sSFR greater than 10$^{-11}$ yr$^{-1}$. Overall, these dwarf galaxies have an $i$-band median apparent magnitude of $\overline{i}\sim 22$ mag and 99\% have $i<24$ mag, which should easily be detected in HSC broad-band imaging data. The Initial Dwarf Sample has a median half-light radius (measured from the {\it HST}/ACS F814W catalog of \citealt[][]{Leauthaud:2007}) of 0.6 arcseconds. The median seeing in the HSC $i$-band is also 0.6 arcsec \citep[][]{Aihara:2018}. Therefore, many dwarfs in the Initial Dwarf Sample will be extended objects in HSC catalogs. Most galaxies in our target mass range \merianmass\ will have a disc like bulge or will be entirely bulgeless \citep{Dutton:2009,Fisher:2011} and can be treated as exponential disks \citep{KadoFong:2020,KadoFong:2021}. For simplicity, we use exponential light profiles when simulating dwarf images given different filter designs (see Section~\ref{filter_simulation}).

\subsection{SED library for the Initial Dwarf Sample}\label{SED_lib}

\begin{figure*}
\begin{center}
\includegraphics[width=18cm]{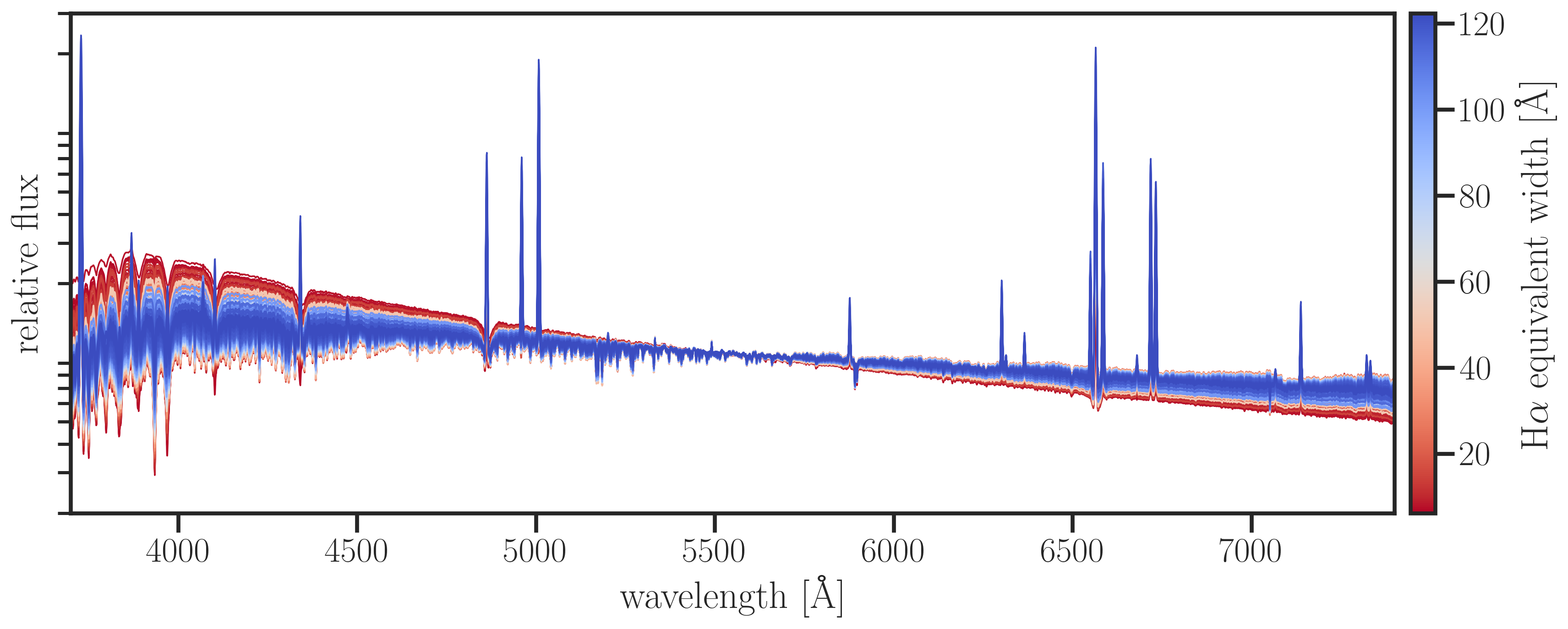}
\caption{Model SEDs (normalized to 5500\AA) for the Initial Dwarf Sample generated with \texttt{Prospector}, color-coded by the $\rm H\alpha$ equivalent width. Every galaxy in this dwarf SED library has a model SED with continuum and emission lines, and are used to predict fluxes for different filter designs.}
\label{spectra_lib}
\end{center}
\end{figure*}

Although COSMOS2015 provides a wealth of photometry and data products from their 30-band SED fitting (e.g., photo-$z$'s, stellar masses, and sSFR), it does not include a best-fit SED with emission lines and continuum for every galaxy, which is crucial for the Merian filter design. Therefore, we run our own SED fitting to the Initial Dwarf Sample to build a dwarf SED library. We now describe the photometric data used for this SED fitting process.

Our analysis of the Initial Dwarf Sample indicates that the photometric data in most of the COSMOS2015 bands have sufficient depth for our goals, with a few exceptions. Only 70\% of the galaxies in the Initial Dwarf Sample have S/N $>$ 3 in GALEX-FUV and GALEX-NUV, while over 95\% of galaxies have S/N $>$ 3 for all other bands in the Subaru COSMOS 20 band survey. In addition, because our dwarf galaxies have relatively small angular sizes, we need the imaging PSF to be uniform through all the bands to avoid blending issues which could affect the SED fitting. For these reasons, we choose to fit the SEDs only using data from MegaCam-u$^*$, HSC-y, VIRCAM-Y/J/H/K$_s$ and the 20 optical filters from the Subaru COSMOS 20 band survey \citep{Taniguchi:2007,Taniguchi:2015}. The Subaru COSMOS 20 band survey contains photometry from 6 broad-bands (B, V, g, r, i$^+$, z$^{++}$), 12 medium-bands (IA427, IA464, IA484, IA505, IA527, IA574, IA624, IA679, IA709, IA738, IA767, IA827), and 2 narrow-bands (NB711, NB816). Note that these medium and narrow band filters were for the now decommissioned Subaru SuprimeCam and therefore cannot be used for Merian. All the bands that we use for SED fitting have S/N$>$3 for more than 90\% of dwarfs in the Initial Dwarf Sample.

We perform SED fitting for the Initial Dwarf Sample using the code \texttt{Prospector} \citep{Johnson:2021}. \texttt{Prospector} is a modular galaxy stellar populations inference code based on Bayesian forward modeling and Monte Carlo sampling of the parameter space using stellar populations generated by the Flexible Stellar Population Synthesis (\texttt{FSPS}) code \citep{Conroy:2009,Conroy:2010}. We use a parametric delayed-$\tau$ star formation history, a \citet{Calzetti:2000} attenuation curve, and free gas-phase and stellar metallicities in the fitting process. We use the MIST stellar evolutionary isochrones and the MILES spectral library in \texttt{FSPS} \citep{Choi:2016,Sanchez-Blazquez:2006} for the stellar population. Nebular continuum and emission lines are included using the pre-compiled \texttt{CloudyFSPS} models in \citet{Byler:2017}. Note that the choice of the stellar isochrones may affect the emission line luminosities due to the treatment of stellar rotation \citep{Byler:2017}. Isochrones which include stellar rotation (e.g. MIST) produce better matches to emission line ratios in diagnostic diagrams such as the Baldwin-Phillips-Terlevich (BPT) diagram \citep{Baldwin:1981}, but they sometimes overestimate the amount of ionized photons \citep{Byler:2017}. There is a possibility that the MIST isochrones overestimate the emission line luminosities in the medium-band filters for some dwarf galaxies, which would affect their observed S/N. However, we defer a more in-depth consideration of this effect for future work.

During the SED fitting process, we fix the redshift to the best fit 30-band photo-$z$ ($\rm z_{COSMOS}$) from the COSMOS2015 catalog. The spec-$z$ is used when available. We use the dynamic nested sampling package \texttt{dynesty} to sample the posteriors \citep{Speagle:2020}. Figure~\ref{spectra_lib} shows the SED library (normalized to 5500\AA) generated by \texttt{Prospector} for the Initial Dwarf Sample, color-coded by the $\rm H\alpha$ equivalent width measured directly from the model SEDs. Each model SED includes both continuum and emission lines and can be utilized to predict fluxes for dwarf galaxies in any filter design. 

\subsection{Predicting Fluxes from SEDs}\label{broadbands}

\begin{figure}
\begin{center}
\includegraphics[width=8cm]{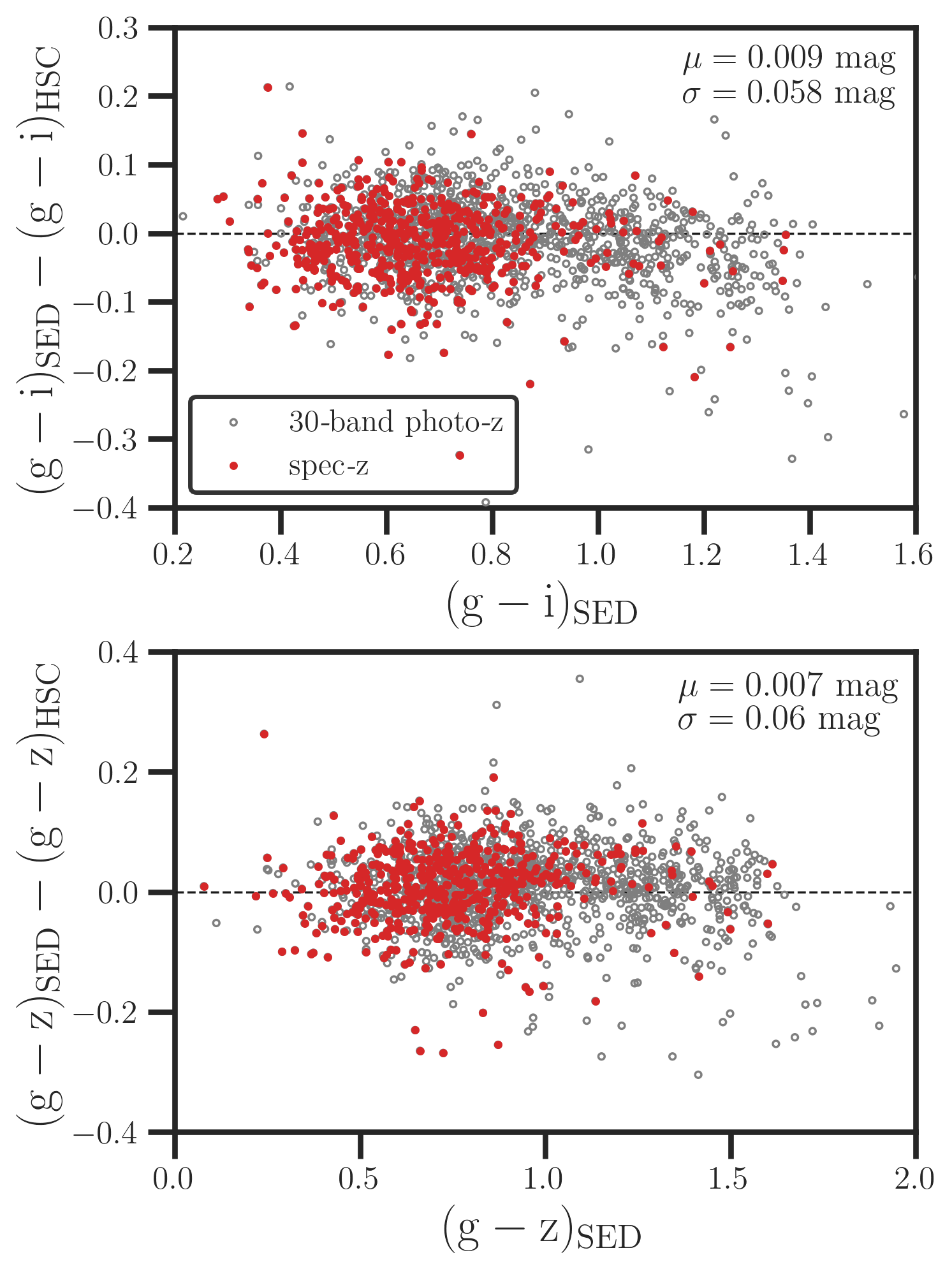}
\caption{Comparison between predicted HSC $g-i$ and $g-z$ colors from our model SEDs and the observed data from HSC-PDR3 for the Initial Dwarf Sample. Galaxies with existing spec-$z$'s are highlighted as red points, while grey open circles are galaxies with a COSMOS 30-band photo-$z$'s. The median $\mu$ and root mean square $\sigma$ of the y-axes for the whole sample (red and grey points) are shown in the upper right corners. Our SED fitting process accurately predicts HSC colors that were not included in our input data vector for the Initial Dwarf Sample.}
\label{SED_HSC}
\end{center}
\end{figure}

To verify that the SEDs we generated for the Initial Dwarf Sample in Section~\ref{SED_lib} produce reasonable fluxes in different filters, we predict fluxes in HSC broad bands by convolving model SEDs with HSC transmission curves and compare them with observed HSC fluxes. Since we care more about the relative colors between different bands than absolute fluxes, we obtain the broad-band PSF-matched aperture photometry (the ``afterburner photometry") with target 1.09 arcsec PSF and 1.5 arcsec apertures taken from HSC-PDR3 database \citep{Aihara:2018}. This is recommended for accurate colors and is the photometry used for HSC photo-$z$ estimates \citep{Tanaka:2018}. 

Figure~\ref{SED_HSC} shows the difference between the HSC $g-i$ (and $g-z$) colors from our model SEDs and the observed data from HSC-PDR3 for the Initial Dwarf Sample. The mean value ($\mu$) of the $g-i$ and $g-z$ difference for the Initial Dwarf Sample is $0.009$ mag and 0.007 mag respectively, with a scatter of $\sigma=0.058$ mag and $\sigma=0.06$ mag. Note that the HSC $g-i$ PSF color has an accuracy of $\mu \sim0.01$ and a precision of $\sigma=0.08$ mag for objects with $i\sim23$ mag \citep{Huang:2018}. The precision of the colors generated from our model SED library is within the HSC photometric precision, which proves that our model SED library provides reliable colors. Galaxies with 30-band photo-$z$'s (grey points) do not have significantly larger scatter than  galaxies with spec-$z$'s (red points), indicating that the precision of the 30-band photo-$z$'s is comparable to spec-$z$'s when generating model SEDs. For the Merian survey, we will use the two Merian medium-band filters along with the five HSC broad-band data to improve the photo-$z$'s. We will use the predicted HSC broad band photometry presented here when characterizing the Merian photo-$z$ performance in Section~\ref{merian_photoz}.

To create a sample of dwarfs at any given redshift,  we convert apparent sizes in arcsec into physical sizes in kpc, and observed SEDs into rest-frame SEDs, assuming no size and SED evolution below $z<0.2$. We then use the full Initial Dwarf Sample to create mock dwarf samples and mock dwarf fluxes for different filter designs. 

\section{Filter design methodology}\label{methodology}
 In this section, we discuss the method used to design the filter set for the Merian survey. The rational for using narrow/medium-band filters is to capture specific spectral features (i.e. emission lines for our target dwarf galaxies) to improve the precision of photo-$z$'s. Since most of the dwarfs in our target mass range are star-forming (as shown in Figure~\ref{dwarf_hist}), they will have strong emission lines. $\rm H\alpha$ and $\rm [OIII]\lambda 5007$ (hereafter $\rm [OIII]$) are often the strongest emission lines for galaxies in the Initial Dwarf Sample. Although $\rm [OII]\lambda 3725$ and $\rm [OII]\lambda 3727$ are also strong emission lines in the Initial Dwarf Sample, their wavelengths are below 4500\AA ~at $z<0.2$. Since DECam is relatively inefficient below 4500\AA\, we only focus on $\rm H\alpha$ and $\rm [OIII]$ for the filter design. We start with the optimization of the $\rm H\alpha$ filter, and then design the $\rm [OIII]$ filter (which helps remove interlopers at higher redshfits, see details in Section~\ref{mock_merian_photometry}) based on the corresponding redshift range given by the optimized $\rm H\alpha$ filter. We optimize the $\rm H\alpha$ filter central wavelength ($\lambda_c$) and full width at half maximum FWHM ($\Delta\lambda$) (that defines the filter bandpass) following the optimization criteria and method described below. 
 

\subsection{Optimization Criteria}\label{optimization_Criteria}

The primary goal of the Merian survey is to build a large photometric sample of dwarf galaxies that can be used to measure the dark matter halos of dwarfs using galaxy-galaxy lensing. For a fixed observing time, there is a trade-off between survey volume and survey depth. More specifically, we design the filter set based on the optimization of two competing requirements:

\begin{itemize}
\item \textbf{Sample Size: we aim to obtain a sample of dwarf galaxies in the mass range \merianmass\ that is large enough to achieve a weak lensing S/N from HSC-SSP greater than 30 within $r<0.5$ Mpc.}. The S/N of the lensing signal depends directly on the number of dwarf (lens) galaxies. A wider redshift range (i.e. larger volume) contains more galaxies. For a specific emission line, this means wider wavelength range (larger filter $\Delta\lambda$). A larger $\lambda_c$ corresponds to a higher central redshift which also yields a larger volume. Thus, this requirement prefers a wider filter $\Delta\lambda$ and larger $\lambda_c$.

\item \textbf{Sample Completeness: we want to ensure that the detection completeness of star-forming dwarf galaxies in the target mass range achieves at least 80$\%$ in the Merian filters.} This requirement prefers a narrower filter $\Delta\lambda$ to ensure that the emission line strength can be distinguished from the continuum. A larger $\lambda_c$, i.e. higher redshift for a given emission line, results in fainter apparent magnitude, which is harder to detect. Therefore this requirement also prefers low $\lambda_c$. In addition, it is important to take into account sky emission lines at specific wavelengths. In general, sky lines become stronger above 8000\AA. Therefore, this requirement prefers filters with narrow $\Delta\lambda$ and central wavelengths below 8000\AA. 

Our final design represents a balance between these competing requirements.

\end{itemize}

\subsection{Lensing Signal-to-Noise Ratio}\label{Lensing_SN}

To evaluate the weak gravitational lensing performance of different filter designs we need to predict the uncertainties on the galaxy-galaxy lensing signal for dwarf galaxies with survey parameters corresponding to a given filter design. The galaxy-galaxy lensing signal that we predict is the excess surface mass density $\Delta\Sigma$, which is defined as:

\begin{equation}
    \Delta\Sigma(r) \equiv \overline{\Sigma}(<r) - \overline{\Sigma}(r) = \Sigma_{c} \times \gamma_t(r),
\end{equation}
where $\overline{\Sigma}(<r)$ is the mean surface density within proper radius r, $\overline{\Sigma}(r)$ is the azimuthally averaged surface density at radius r, $\Sigma_{c}$ is the critical surface mass density, and $\gamma_t(r)$ is the tangentially projected shear \citep[e.g.][]{Miralda-Escude:1991,Wilson:2001,Leauthaud:2012}. The lensing S/N depends on the number of lens galaxies (in our case, these are the target dwarf galaxies) and the number density of source galaxies. The number of lens galaxies is determined by the area and volume of the medium band survey. The redshift range of the survey is set by the wavelength range of the filter. 

We use the HSC-SSP wide layer for the basis for our predictions. To predict the number of lens galaxies, we assume a survey area that can take on a  maximum value of 1000 deg$^2$. We use the COSMOS2015 stellar mass function to compute the number density of dwarf galaxies given redshift range within the mass range \merianmass. The combination of the galaxy number density and the survey volume gives the number of lens galaxies. For source galaxies, we assume the HSC S16A source density and a mean redshift of $\langle z_s \rangle = 0.81$ \citep[e.g.][]{Hikage:2019}.

We compute the predicted lensing uncertainties for $\Delta\Sigma$ using the same methodology described in \citet{Singh:2017}. Here we briefly summarize the salient features of this computation and refer the reader to \citet{Singh:2017} for further details. The Gaussian covariance for the galaxy–galaxy lensing measurement $\Delta\Sigma$ within a projected separation $r$ is given by:

\begin{displaymath}
  {\rm Cov}(\ds(r),\ds(r'))\approx  \frac{\Sigma_c^2(\chi_s,\chi_g)}{V_W} \int dk_\perp k_\perp J_2(k_\perp r)
\end{displaymath}
\begin{equation}
\label{cov}
 J_2(k_\perp r')\left[\left(P_{gg}(k_\perp)+\frac{1}{n_g}\right)\left(P_{\kappa\kappa}(k_\perp)+\frac{\sigma^2}{n_s}\right)+P_{g\kappa}^2(k_\perp)\right],
\end{equation}

\noindent where $\Sigma_c$ is the critical surface density, $\chi_g$ and $\chi_s$ are the comoving distances to lens and source galaxies, $V_W$ is the physical volume of the survey, $k_\perp$ is the wave vector perpendicular to the line-of-sight, $J_2$ is the second-order spherical Bessel function, $P_{gg}$ is the lens galaxy power spectrum, $\sigma$ is the shape noise, $n_s$ is the source galaxy density, $n_g$ is the lens galaxy density, $\frac{\sigma^2}{n_s}$ is the shape noise term, and $\frac{1}{n_g}$ is the lens galaxy shot noise term. For the power spectrum, we use the {\tt HaloFit} nonlinear power spectrum \citep{Smith:2003,Takahashi:2012}. For the lens galaxy power spectrum $P_{gg}$, we use linear galaxy bias with the non-linear matter power spectrum. The galaxy-matter power spectrum ($\Sigma_c^2P_{g\kappa}$) is obtained by direct inverse Hankel transform of $\ds$. The convergence power spectrum, $P_{\kappa\kappa}$ in units of $P(k)$ is given by:
\begin{equation}
    P_{\kappa\kappa}(k_\perp)=\int_0^{\chi_s}d\chi \frac{\overline{\rho}^2}{\Sigma_c^2(\chi_s,\chi)}P_{mm}(k_\perp\frac{\chi_g}{\chi}),
\end{equation}
where $\overline{\rho}$ is the mean density and $P_{mm}$ is the matter power spectrum.

These $\Delta\Sigma$ errors include all the terms needed to describe the Gaussian covariance. We do not account for the effects due to selection functions or complicated survey masks. We also ignore contributions from the connected covariance which includes super sample covariance, as well as the trispectrum between galaxies and shear. These ignored terms will result in an overestimated lensing S/N by up to $\sim$25\%, hence we apply a $\sim$25\% reduction in our lensing S/N estimates. The exact normalization of this calculation is not important for this paper since the filter design uses relative comparisons between S/N estimates. For further details, please see \citet[][]{Singh:2017, Leauthaud:2020}.

\subsection{Simulating Filter Performance}
\label{filter_simulation}

We now describe our method for predicting the performance of various \Ha\ filter designs and optimizing the filter central wavelength ($\lambda_c$) and full width at half maximum FWHM ($\Delta\lambda$) based on the optimization criteria presented in Section~\ref{optimization_Criteria}. As discussed in Section~\ref{Initial_Dwarf_Sample}, some dwarf galaxies in the Initial Dwarf Sample are extended sources in HSC. In order to more realistically predict the fluxes and uncertainties for dwarf galaxies in different filter designs, we generate mock images to run the filter optimization. We generate dwarf mock images as follows.

\begin{itemize}

\item For each \Ha\ filter design (a combination of $\lambda_c$ and $\Delta\lambda$), we re-assign physical properties for every dwarf galaxy in the Initial Dwarf Sample. We consider the redshift range corresponding to a given filter choice. The redshift range is derived from the wavelength range given $\lambda_c$ and $\Delta\lambda$ for the \Ha\ emission line. We then assign a random redshift to every galaxy in the Initial Dwarf Sample drawn uniformly from the redshift range of a given filter choice. We convert physical sizes in kpc to apparent sizes in arcsec using size estimates from the COSMOS ACS catalog. We then convert the SED from the rest-frame to the observed-frame using this newly assigned redshift and then compute fluxes in the HSC broad bands and the Merian \Ha\ band. 

\item We create dwarf galaxy images assuming that the light profiles of our target dwarf galaxies follow circular exponential profiles with a face-on orientation. The choice of orientation is a conservative assumption since it corresponds to the lowest S/N of fluxes due to minimum surface brightness. We create mock postage stamp images for these galaxies with a given exposure time and the instrumental parameters of DECam using the galaxy image simulation tool {\tt Galsim} \citep{Rowe:2015}. We assume a fixed 1.0 arcsec seeing and dark time condition. The sky background is added using a model optical sky spectrum for CTIO. Mock fluxes and S/N values were checked against the Exposure Time Calculator (ETC) provided by DECam.

\item We generate dwarf galaxy images with and without emission lines (i.e. continuum only) for each filter design, assuming the same spatial distributions for the continuum and emission line regions. We then use the photometry code {\tt SEP} \citep{Barbary:2015} to measure the total fluxes of galaxies. We use {\tt FLUX\_AUTO} from {\tt SEP} as well as the associated flux uncertainties ({\tt FLUXERR\_AUTO}). The flux with emission lines is noted $\rm Flux_{EL}$ and the flux without emission lines is noted $\rm Flux_{CONT}$.

\end{itemize}

At this stage, we assume that galaxy centroids are known, since the position of galaxies can be well-determined from the deeper broad-band survey data. Indeed, Merian aims to take advantage of the deeper HSC photometry to perform forced photometry on the shallower medium band images. We now need a criterion to determine if the emission line of a dwarf galaxy has been detected in the medium band filters. We consider an emission line to be detected if $\rm Flux_{EL}-Flux_{CONT}>5\times Fluxerr_{EL}$ where $\rm Fluxerr_{EL}$ is the {\tt FLUXERR\_AUTO} for images with emission lines. 

We run image simulations for different filter designs with $\lambda_c$ corresponding to \Ha\ at $0.02<z<0.2$ and $\Delta\lambda$ values between 100\AA\ and 400\AA, and starting with 600-second exposures on DECam. The upper limit of the total survey area is set to be $\rm 1000\ deg^{2}$ given the size of the HSC-SSP wide layer. During the optimization, if the detected dwarf fraction is below 80\%, we increase the exposure time by 300s. Longer exposure times result in smaller survey areas at a fixed total number of nights. Note that the filter design does not depend on the total number of nights, only the predicted lensing S/N depends on the number of nights. We adopt 24 full nights for the \Ha\ filter design, which is the actual number of nights allocated to the \Ha\ filter for Merian. In practice, Merian has 60 full nights in total, where 36 nights are dedicated to the $\rm [OIII]$ filter (see Section~\ref{final_filter_design} for details).

We use the stellar mass function from \citet{Davidzon:2017} to compute the number of dwarf galaxies with \merianmass\ in the redshift range corresponding to each filter design. We multiply the total number of galaxies by 80\% to roughly account for the expected sky area loss due to bright stars \citep[e.g.,][]{Leauthaud:2007, Heymans:2012,Coupon:2018}. For each filter design, we compute the total number of detected dwarf galaxies, and the survey area assuming 24 full nights on DECam for the $\rm H\alpha$ filter. The lensing S/N is then computed using the method described in section~\ref{Lensing_SN} based on the corresponding redshift range and the total number of dwarf galaxies (which serve as lens galaxies). 

\begin{figure*}
\begin{center}
\includegraphics[width=18cm]{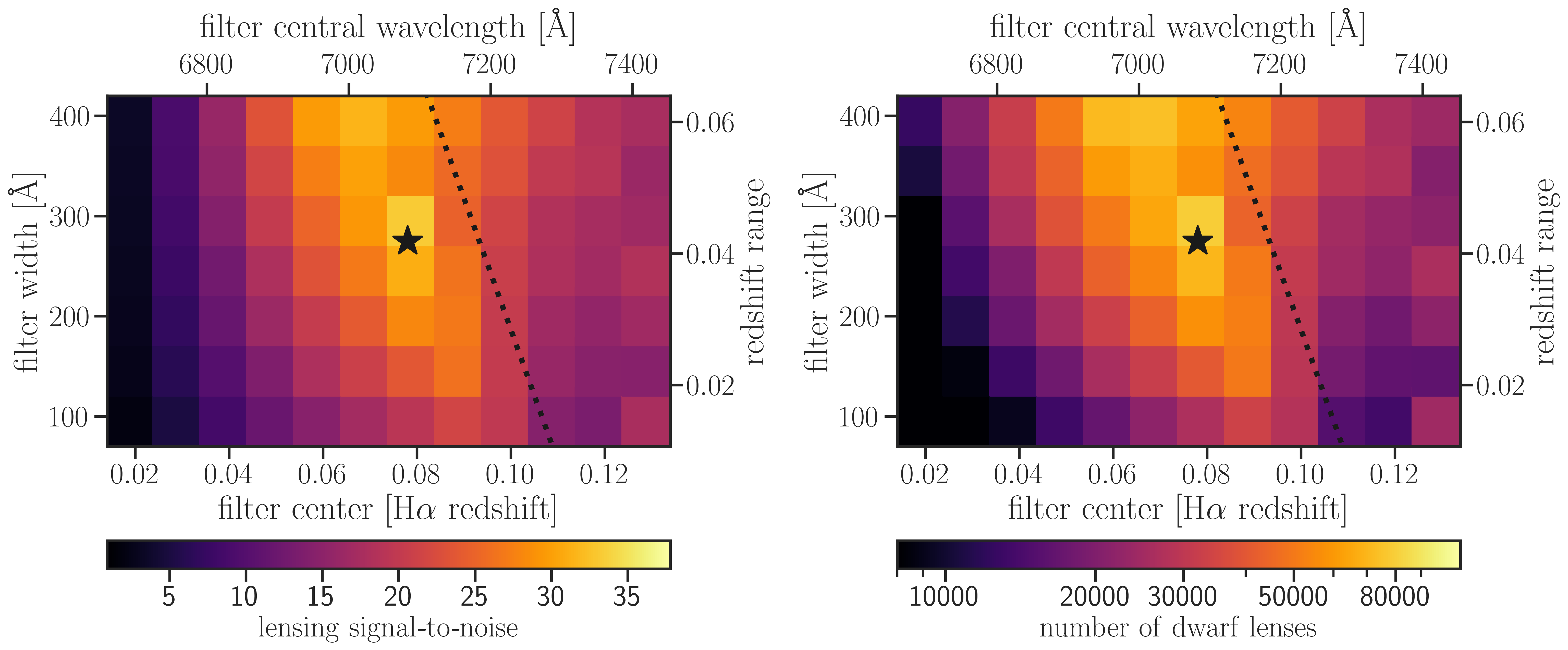}
\caption{Filter width vs. filter central wavelength for different designs of the $\rm H\alpha$ filter. Upper and lower x-axes show the filter central wavelength in \AA\ and the corresponding redshift for $\rm H\alpha$. Left and right y-axes show the filter width in wavelength and the corresponding redshift range. Colors in the left panel represent the predicted weak lensing S/N at $r<0.5$ Mpc given a fixed observing time of 24 full nights on DECam (for $\rm H\alpha$). The right panel shows the expected number of dwarf galaxies (lenses) for each filter design. The dotted diagonal lines indicate the location of the strong sky emission line at 5580\AA\ for a matched $\rm [OIII]$ filter. The star shows our final choice for the N708 $\rm H\alpha$ filter which is a compromise between predicted lensing S/N, the strength of excess fluxes between medium-band and broad-band filters, and the desire to avoid strong skylines in the matched $\rm [OIII]$ filter.}
\label{lensing_SN_HSC}
\end{center}
\end{figure*}

\begin{figure}
\begin{center}
\includegraphics[width=8cm]{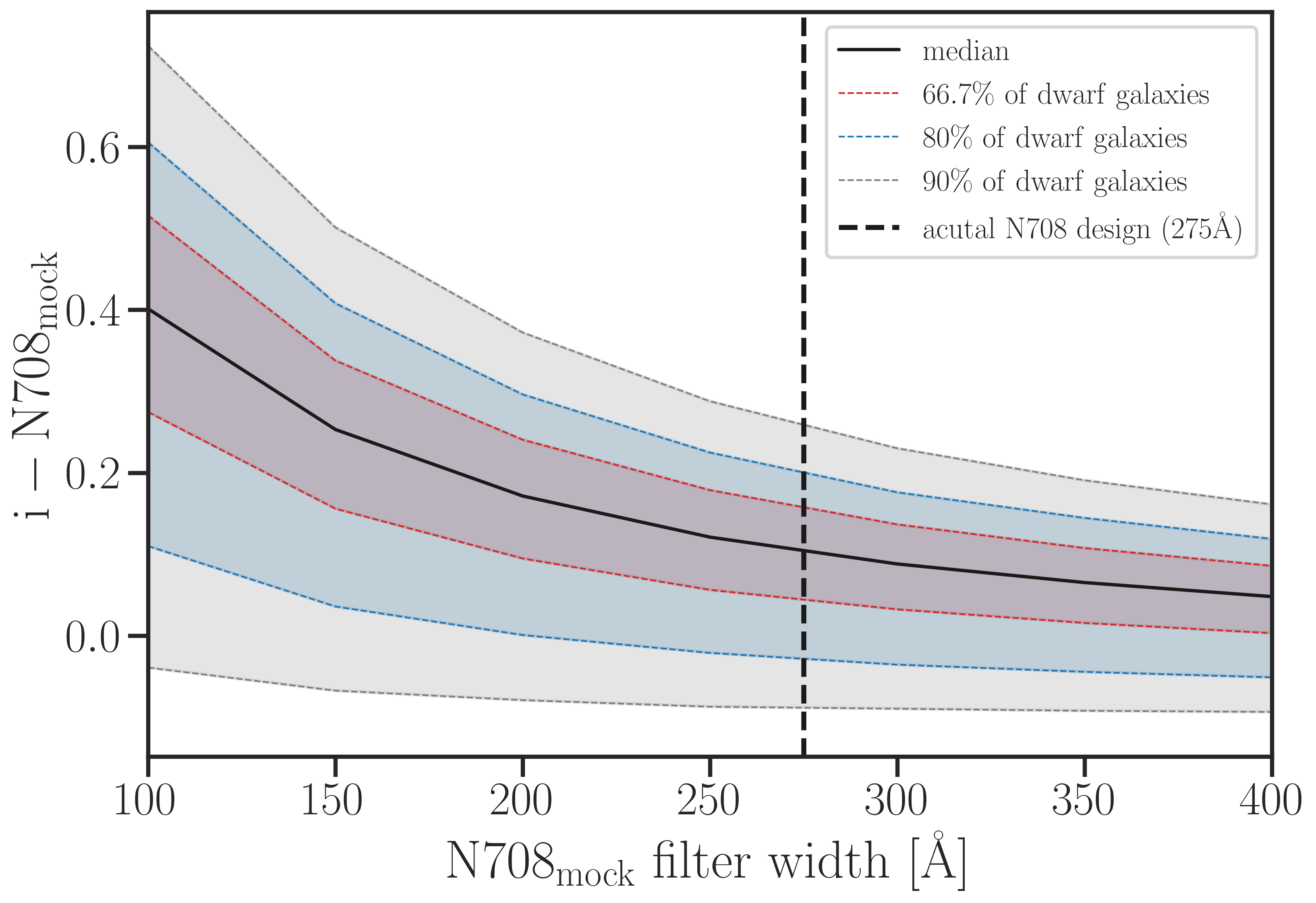}
\caption{Mock colors between the HSC $i$-band and filters with central wavelength set to 7080\AA\ and varying filter widths for the Initial Dwarf Sample shifted within each filter design. The red, blue, and grey shaded areas show the 66.7\%, 80\%, and 90\% width of the magnitude difference distribution, while the black curve shows the median. The black vertical dashed line indicates the final Merian N708 filter width (275\AA).}
\label{deltamag}
\end{center}
\end{figure}

\begin{figure*}
\begin{center}
\includegraphics[width=16cm]{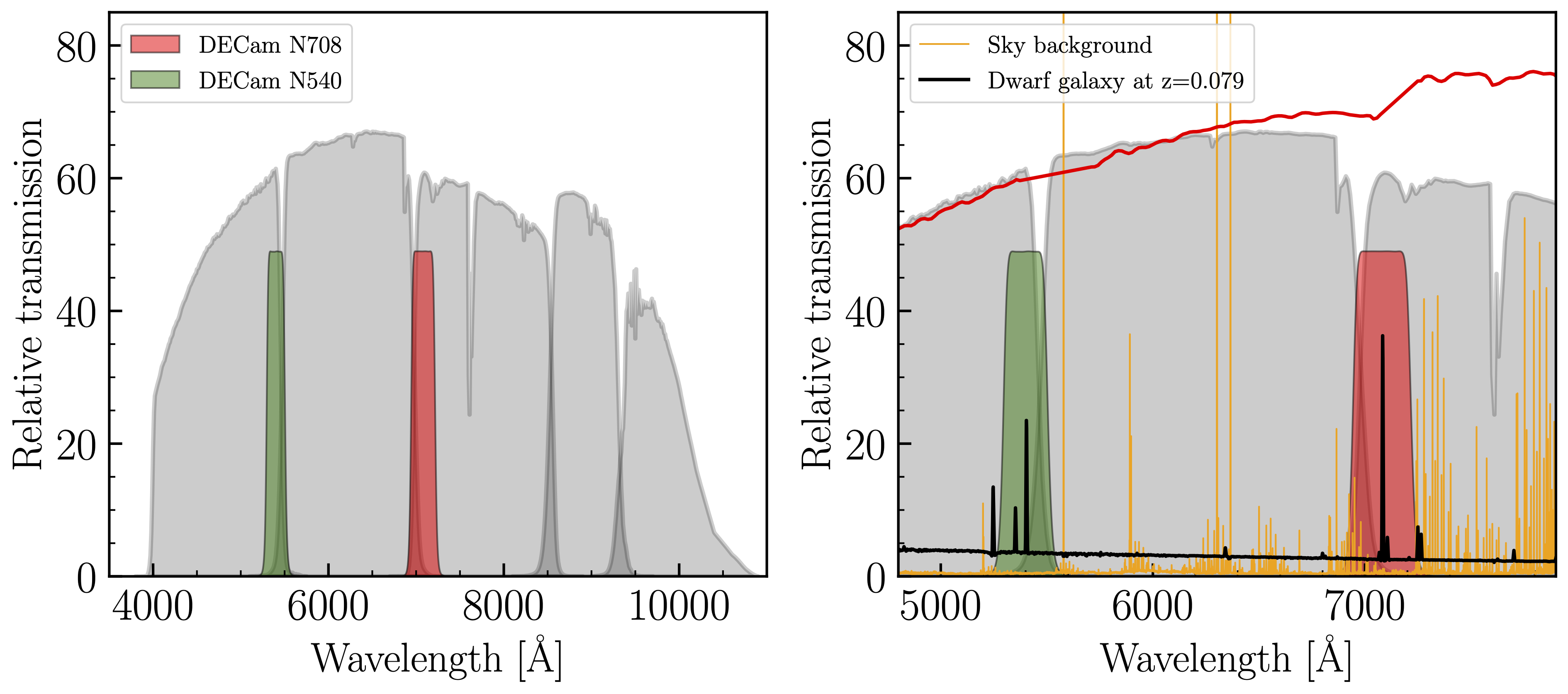}
\caption{Left panel: Relative transmission curves for DECam N708 and N540 compared to HSC $grizy$ broad-band filters (grey shaded area). Right panel: Zoom-in region for N708 and N540, with sky emission (yellow) and an example dwarf galaxy spectrum at $z=0.079$ (black). The red solid curve indicates the DECam response curve in this wavelength range.}
\label{filter_transmission}
\end{center}
\end{figure*}

\begin{figure*}
\begin{center}
\includegraphics[width=15cm]{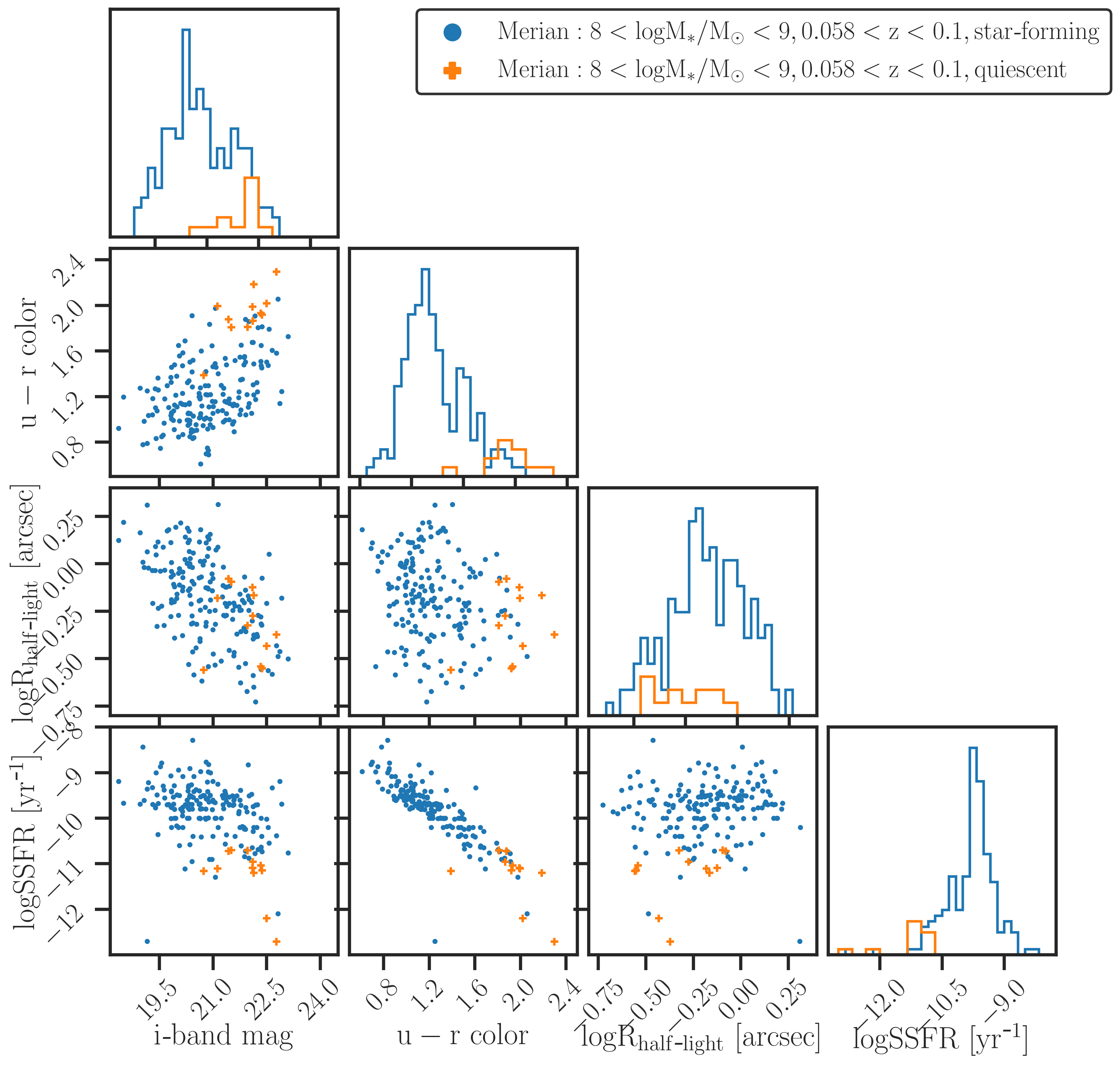}
\caption{Distributions of four basic properties ($i$-band magnitude, $u-r$ color, half-light radius, and specific star-formation rate) for the target dwarf galaxies from the Initial Dwarf Sample corresponding to the Merian filter set (N708 and N540). Blue points (histograms) and orange crosses (histograms) correspond respectively to star-forming and quiescent dwarf galaxies in the Merian target mass and redshift range (\merianmass,\ $\rm 0.058<z_{COSMOS}<0.1$) according to the classification in the COSMOS2015 catalog.}
\label{dwarf_hist_merian}
\end{center}
\end{figure*}

\begin{figure*}
\begin{center}
\includegraphics[width=18cm]{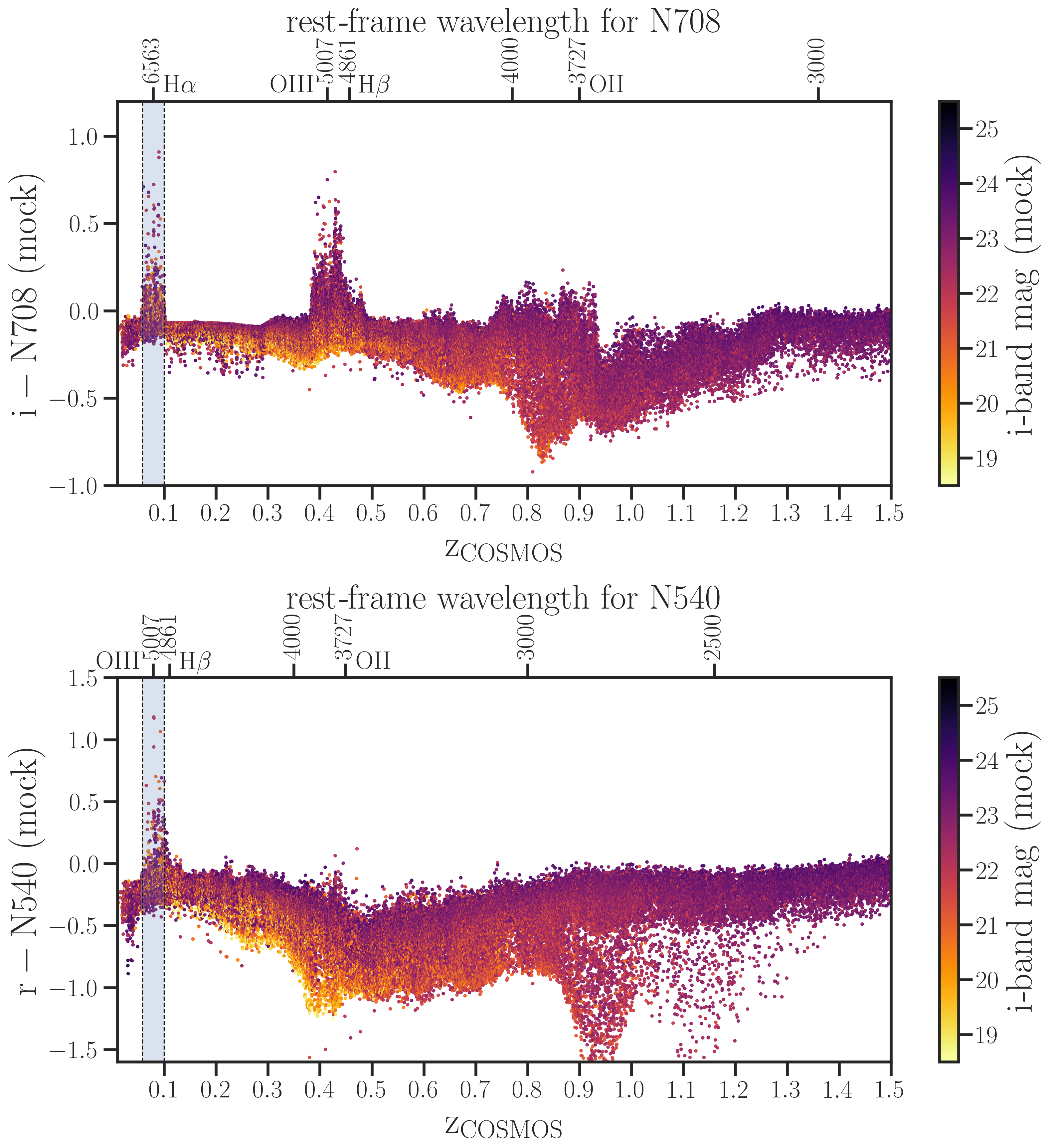}
\caption{Mock noise-free \iha\ and \roiii\ color vs. redshift color-coded by mock $i$-band magnitude for galaxies (all masses) in the COSMOS2015 catalog. Mock photometry in $i$-band, N708 and N540 are computed from model SEDs generated following the same procedure as our dwarf SED library for the Initial Dwarf Sample (see details in Section~\ref{SED_lib}). The upper x-axis indicates the corresponding rest-frame wavelengths for N708 and N540 (i.e. 7080\AA/(1+z) and 5400\AA/(1+z)), including the rest-frame wavelengths for $\rm H\alpha$, $\rm [OIII]$, $\rm H\beta$, and $\rm [OII]$. The shaded areas show the target Merian redshift range $0.058<z<0.1$.}
\label{flux_excess_mock}
\end{center}
\end{figure*}

\begin{table}
 \caption{Characteristics of the final Merian N708 and N540 filters and Merian survey design.}
 \label{table:filter_info}
 \begin{tabular}{|p{3.5cm}|p{1.8cm}|p{1.8cm}|}
  \hline
  & N708 ($\rm H\alpha$) & N540 ($\rm [OIII]$)  \\
  \hline
  Central wavelength $\lambda_c$ (\AA) &  7080      & 5400 \\
  Filter width $\Delta\lambda$ (\AA) &   275     & 210  \\
  Exposure time (min) &  40 & 60   \\
  N$_{\rm nights}$  &   24 & 36 \\
  \end{tabular}
 \begin{tabular}{|p{3.5cm}|p{4cm}|}
  \hline
  Redshift range    & $0.058 < z < 0.10$ \\
  Mass range    & \merianmass \\
  Survey area (deg$^2$)   &   864       \\
  Expected number of dwarfs & 85,447 \\
  Lensing S/N at $r<0.5$ Mpc  & 32 \\
  Lensing S/N at $r<1.0$ Mpc  & 90  \\
  \hline
  \end{tabular}
\label{filter_design}
\end{table}

\section{Filter Optimization Results}\label{results}

\begin{figure*}
\begin{center}
\includegraphics[width=16cm]{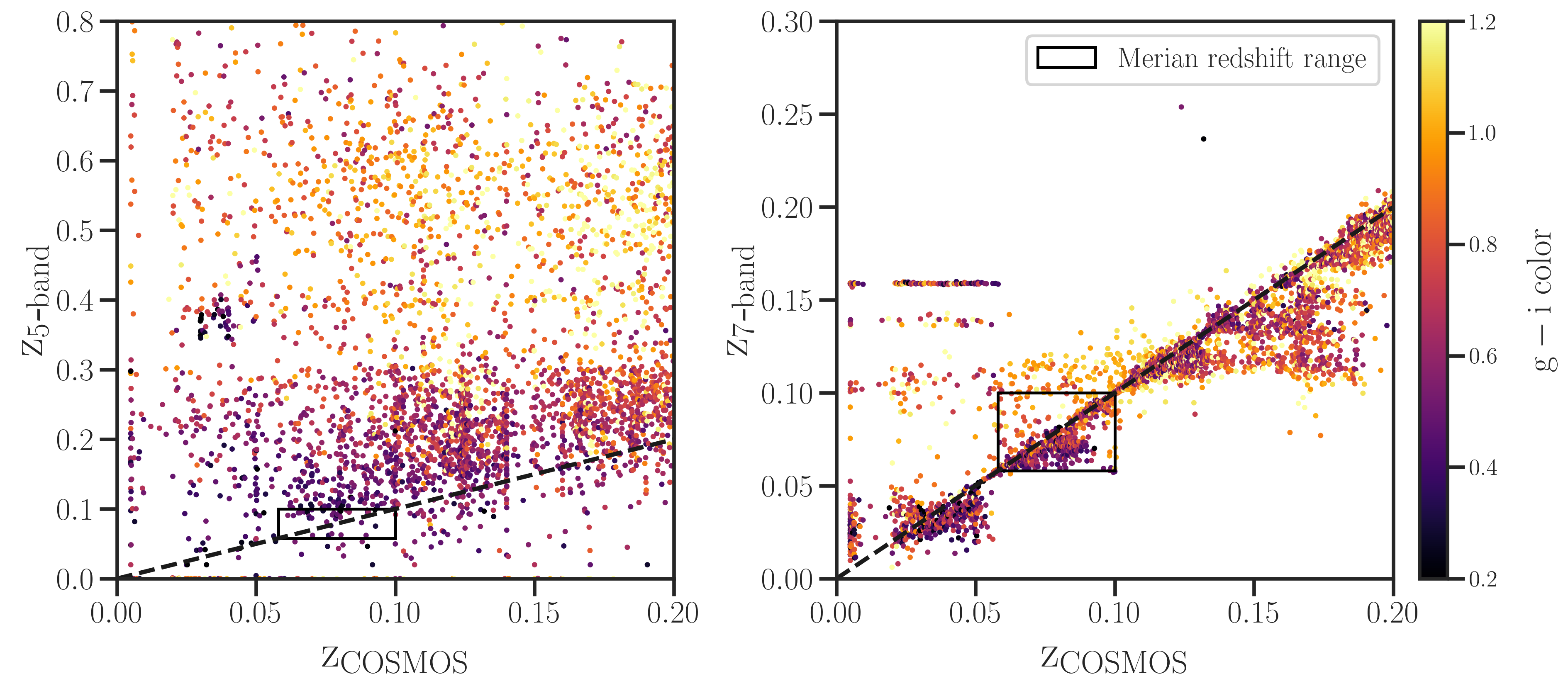}
\caption{HSC 5-band photo-$z$'s (left panel) and mock 7-band photo-$z$'s (right panel) vs. $\rm z_{COSMOS}$, color-coded by $g-i$ color. Black dashed lines show $\rm z_{phot} = \rm z_{COSMOS}$. The black boxes indicate our target redshift range $0.058<z<0.1$. Note that we only show galaxies with HSC 7-band photo-$z$ lower than 0.3 in the right panel since there are no outliers above this cut, while the scatter of HSC 5-band photo-$z$'s extends to higher redshifts. The 5-band HSC broad-band selection has a photo-$z$ accuracy of $\sigma_{\Delta z/(1+z)}\sim 0.5$ and an outlier fraction of $\eta=60\%$, with 48\% completeness and 12\% purity at $z<0.1$. The 7-band selection has a photo-$z$ accuracy of $\sigma_{\Delta z/(1+z)}\sim 0.01$ and an outlier fraction of $\eta=2.8\%$, with 89\% completeness and 90\% purity at $z<0.1$.}
\label{photoz_comparison}
\end{center}
\end{figure*}

\begin{figure}
\begin{center}
\includegraphics[width=8cm]{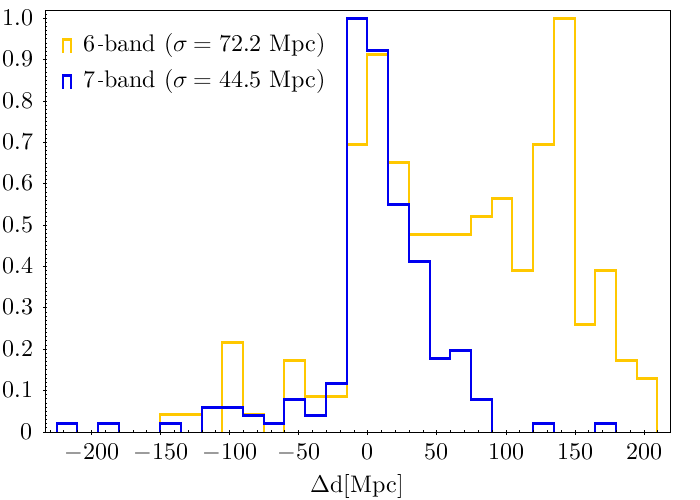}
\caption{Predicted line-of-sight distance localization precision $\rm \Delta d$ for Merian 6 and 7-band photo-$z$'s using idealized photometry. The predicted standard deviation of $\rm \Delta d$ for Merian 7-band photo-$z$ (blue) and 6-band photo-$z$ (yellow) is 44.5 Mpc and 72.2 Mpc, respectively.}
\label{delta_d}
\end{center}
\end{figure}

In this section, we present the optimization of the \Ha\ filter following the methodology described in Section~\ref{methodology}. We then explain the design process for the matched $\rm [OIII]$ filter and for the final joint design. To evaluate the dwarf identification success rate in the target redshift range, we compute mock photometry in the final Merian \Ha\ and $\rm [OIII]$ filter designs. We use mock photometry for  Merian (medium band) and HSC (broad band) to predict photo-$z$ performance.

\subsection{Predicted lensing S/N and Merian filter design}
\label{final_filter_design}

We now present the results of the filter optimization. Figure \ref{lensing_SN_HSC} displays $\Delta\lambda$ vs. $\lambda_c$ where $\lambda_c$ has been converted into the corresponding redshift $z_c$ for $\rm H\alpha$. Colors in the left panel indicate the predicted lensing S/N within 0.5 Mpc. Colors in the right panel indicate the  total number of detected dwarf galaxies for each filter design. The predicted lensing S/N peaks at S/N $\sim$ 32 with a filter width of $\Delta\lambda \sim$ 300\AA\ and a filter center at $z_c\sim0.08$ in units of \Ha\ redshift, assuming a fixed observing time of 24 full nights on DECam.

In general, the lensing S/N decreases above $z_c\sim 0.1$. This is because higher redshift galaxies need longer exposure times which results in a smaller survey area, volume, and therefore also a smaller lens sample. We run the optimization for \Ha\ between $0.02<z_c<0.2$ but Figure \ref{lensing_SN_HSC} only shows the results for $0.02<z_c<0.13$ because the lensing S/N continues to decrease at $z_c>0.13$. The lensing S/N also decreases below $z_c\sim 0.05$. This is the result of the fixed area of the HSC-SSP wide layer. Indeed, while lower redshift dwarf galaxies are easier to detect with shorter exposure times, the volume becomes limited by the fact that HSC has a fixed area of $\rm \sim 1000\ deg^{2}$. Note that a similar optimization applied to the Rubin LSST survey might prefer lower redshifts due to the wider area that will be available from LSST \citep[][]{Ivezi:2019}. 

In the optimization procedure, for simplicity, we do not measure photo-z's for every filter design. Instead, we measure the fluxes for mock dwarfs with and without emission lines (i.e. continuum only) and characterize the strength of the detection based on the difference between the two. However, in practice we do not have the fluxes for the continuum underlying \Ha, only the HSC broad-band photometry. The closest broad-band filters would inevitably also include emission lines, which means we do not have a clean continuum to subtract. Therefore, we rely on the photo-$z$/SED fitting procedure to produce a reliable model of \Ha\ continuum  to fit the \Ha\ emission correctly and also to determine redshifts. The reliability of the model continuum from this fitting process depends on the photometric precision of Merian and HSC, which is difficult to predict without a comprehensive understanding of the real data. In general, it is not easy to achieve accurate, unbiased, and consistent photometry at the 0.1 mag level for galaxies at $i\sim 23$ in large imaging surveys \citep{Huang:2018}. Therefore, in practice, we need the magnitude difference between our medium-band and broad-band filter to be large enough to not be affected by systematics in the photometry.

Figure \ref{deltamag} shows the mock magnitude difference between $i$-band and a series of filters with central wavelength set to 7080\AA \ and different filter widths for the Initial Dwarf Sample. We choose the central wavelength to be 7080\AA \ for Figure \ref{deltamag} because our final \Ha\ filter design is centered at 7080\AA. The red, blue, and grey shaded areas in Figure \ref{deltamag} show the 66.7\%, 80\%, and 90\% width of the magnitude difference distribution, while the black curve shows the median. Narrower filters include less continuum and hence correspond to larger magnitude differences. As shown in Figure \ref{deltamag}, the filter design corresponding to the lensing S/N peak ($\Delta\lambda \sim$ 300\AA, $z_c\sim0.08$) in Figure \ref{lensing_SN_HSC} has a fairly large $i-\rm N708_{mock}$ color. 90\% of the dwarfs in this filter have $i-\rm N708_{mock}>-0.1$ and 50\% of the dwarfs have $i-\rm N708_{mock}>0.1$. As expected, the $i-\rm N708_{mock}$ color continues to decrease as the filter width becomes wider because the continuum contributes more to the total flux, thereby decreasing the detection of the \Ha\ emission line. Our final filter design for N708 ($\lambda_c = 7080$\AA, $\Delta\lambda = 275$\AA) is shown as a vertical dashed line in Figure \ref{deltamag} and with black stars in Figure~\ref{lensing_SN_HSC}. In practice, we adjusted the filter width of the final Merian \Ha\ filter design from 300\AA\ to 275\AA\ because of the $\rm [OIII]$ matched filter. This is now discussed below.

The purpose of a second Merian filter (targeting $\rm [OIII]$) is to reduce contamination and improve Merian redshifts. The combination of the $\rm H\alpha$ and $\rm [OIII]$ matched filter set will help to limit interlopers from higher redshifts. For example, with a single filter design, the $\rm [OIII]$ emission line moving into the $\rm H\alpha$ filter will cause contamination from galaxies at $z\sim 0.4$ (see Section~\ref{mock_merian_photometry} for more discussion). The wavelength range of the $\rm [OIII]$ filter is set to match the redshift range probed by the $\rm H\alpha$ filter. The predicted S/N of the lensing peaks for $\rm H\alpha$ filter designs with $z_c\sim 0.08$, $\Delta\lambda \sim 300$\AA\ (6939\AA\ $< \lambda <$ 7239\AA), corresponding to $0.057<z_c<0.103$. The corresponding wavelength range for the $\rm [OIII]$ filter is 5293\AA\ $< \lambda <$ 5522\AA. However, we wish to avoid a strong sky emission line at 5580\AA. Although our filters will have rough top-hat transmission curves, wings in the transmission curves can extend the wavelength range ($\pm$ 10\AA). In addition, transmission curves can also shift ($\sim$ 50\AA) between the filter center and the filter edge due to different angles of incidence. This will be shown in Section~\ref{filter_construction} with real Merian filter transmission curves. Therefore, we opt for a conservative design and reduce the wavelength/redshift range and shift $\lambda_c$  slightly compared to the optimum value from Figure \ref{lensing_SN_HSC}. Table \ref{filter_design} summarizes the final design and characteristics of the Merian dual filter set. The Merian $\rm H\alpha$ (N708) filter has $\lambda_c = 7080$\AA\ and $\Delta\lambda = 275$\AA. The Merian $\rm [OIII]$ (N540) filter has $\lambda_c = 5400$\AA\ and $\Delta\lambda = 210$\AA. The optimized exposure times for N708 and N540 are 40 min and 60 min, respectively. 

The exposure time for the $\rm [OIII]$ filter is determined by the same image simulator presented in Section~\ref{filter_simulation}. We assume the same seeing and moon phase as in Section~\ref{filter_simulation} and only change the filter transmission curve. On average, the $\rm [OIII]$ emission line is slightly weaker than \Ha\ (based on our dwarf SED library). Furthermore, the DECam response efficiency at 5400\AA\ is lower than 7080\AA\ (see the red solid curve in Figure \ref{filter_transmission}). For these two reasons, we need 60 min exposures for the $\rm [OIII]$ filter to achieve a similar depth and survey area compared to N708 (40 min). Both exposure times are set to achieve a 80\% completeness limit in our image simulations, as described in Section~\ref{filter_simulation}. In total, Merian has 60 full nights dedicated to the Merian wide layer survey. These 60 nights are broken into 24 full nights for the \Ha\ filter and 36 full nights for the $\rm [OIII]$ filter. The detailed Merian survey design will be presented in Danieli et al. in prep.

Based on our simulated estimates (see Section~\ref{filter_simulation}), we will obtain 85,447 dwarf galaxies over 864 deg$^2$ with a total of 60 nights on DECam with Merian. This sample will allow us to measure the weak lensing signal with a lensing S/N = 32 within $r<0.5$ Mpc and lensing S/N = 90 within $r<1.0$ Mpc. The Merian redshift range of $0.058<z<0.1$ is narrow and will allow for accurate measurements of galaxy-galaxy lensing profiles with less smearing in the radial profile compared to samples selected based on photo-$z$'s derived from broad-band photometry. Figure \ref{filter_transmission} shows the final filter transmission curves for N708 and N540 along with the HSC broad-band filters. The right panel of Figure \ref{filter_transmission} shows a zoom-in region of N708/N540 together with the sky spectrum of a mock dwarf galaxy at $z=0.079$. The red solid line indicates the response curve of DECam.

Figure~\ref{dwarf_hist_merian} displays the properties of dwarf galaxies from the Initial Dwarf Sample that are in the final Merian redshift and mass range (\merianmass\ and $\rm 0.058<z_{COSMOS}<0.1$) . Final Merian target galaxies have similar basic properties (color, size, sSFR) as the Initial Dwarf Sample, but have brighter apparent magnitudes ($\sim$1 mag brighter) because they are at lower redshifts. Roughly 95\% of star-forming dwarfs in the Merian target range have $i<23.3$~mag.

\subsection{Mock Merian Photometry}
\label{mock_merian_photometry}

We now aim to predict the photo-$z$ performance when we combine the HSC photometry with our custom N708 and N540 medium-band filters. The exact photo-$z$ performance will depend on the quality of the photometry, and systematic issues such as deblending, which are not trivial to model in simulations. Hence, the estimates presented here are theoretical scenarios. To estimate photo-$z$ performance, we first generate mock Merian synthetic photometry.

Following the same procedure as in  Section~\ref{SED_lib}, we run SED fitting for all galaxies in our COSMOS2015 sample below $z<0.65$ and compute mock fluxes in N708, N540, and HSC $grizy$ given the best-fit model SEDs. Figure \ref{flux_excess_mock} shows mock \iha\ and \roiii\ color vs. $\rm z_{COSMOS}$ color-coded by mock $i$-band magnitude for all galaxies in this sample. Upper x-axes show the rest-frame wavelength corresponding to the redshifts of $\rm H\alpha$, $\rm [OIII]$, $\rm H\beta$ and $\rm [OII]$ in N708 and N540. Mock N708 (N540) magnitudes are the sum of the fluxes from emission lines and the continuum in N708 (N540), while the $i$-band ($r$-band) magnitude broadly traces the continuum in N708 (N540).

The upper panel of Figure \ref{flux_excess_mock} shows that a simple \iha\ color can be used to select star-forming galaxies between $0.058<z<0.10$. For example, most \Ha\ emitting galaxies have $i-\rm N708 > -0.1$. However, emission lines with shorter wavelengths (e.g. [OIII]) can shift into N708 and will create interlopers from galaxies at higher redshifts. In fact, the number of higher redshift interlopers could outnumber Merian dwarfs, due to the larger cosmological volume at higher redshifts. $\rm [OIII]\lambda 5007$, $\rm [OIII]\lambda 4959$ and $\rm H\beta$ contribute most interlopers at $z\sim0.44$. The lower panel in Figure \ref{flux_excess_mock} suggests that a simple \roiii\ color cut (e.g. $r-\rm N540>-0.2$) could exclude a good fraction of the $z\sim0.44$ interlopers. Removing these interlopers is a key motivation for the $\rm [OIII]$ filter.  

 In our target redshift range $0.058<z<0.10$, the three emission lines $\rm [OIII]\lambda 5007$, $\rm [OIII]\lambda 4959$ and $\rm H\beta$ can shift in and out of N540 from $z=0.058$ to $z=0.10$. Fluxes in N540 will be brighter if they include more than one emission line. Therefore N540 will provide extra redshift information for galaxies in our target redshift range $0.058<z<0.10$. We will show in the next section that photo-$z$'s based on 7-band photometry (N708, N540, and $grizy$) can select dwarf galaxies in the range $0.058<z<0.10$ and may also have some redshift discriminating precision within the target redshift range. 

Note that mock fluxes are generated directly from the best-fit model SED via \texttt{Prospector} and are noise free. In practice, Merian photometry will have systematic errors arising from various steps in the data reduction process and the photometry measurement method, which could impact the redshift precision. However, characterizing photometric uncertainties in Merian is beyond the scope of this paper and will be presented in future work. 

\begin{figure*}
\begin{center}
\includegraphics[width=16cm]{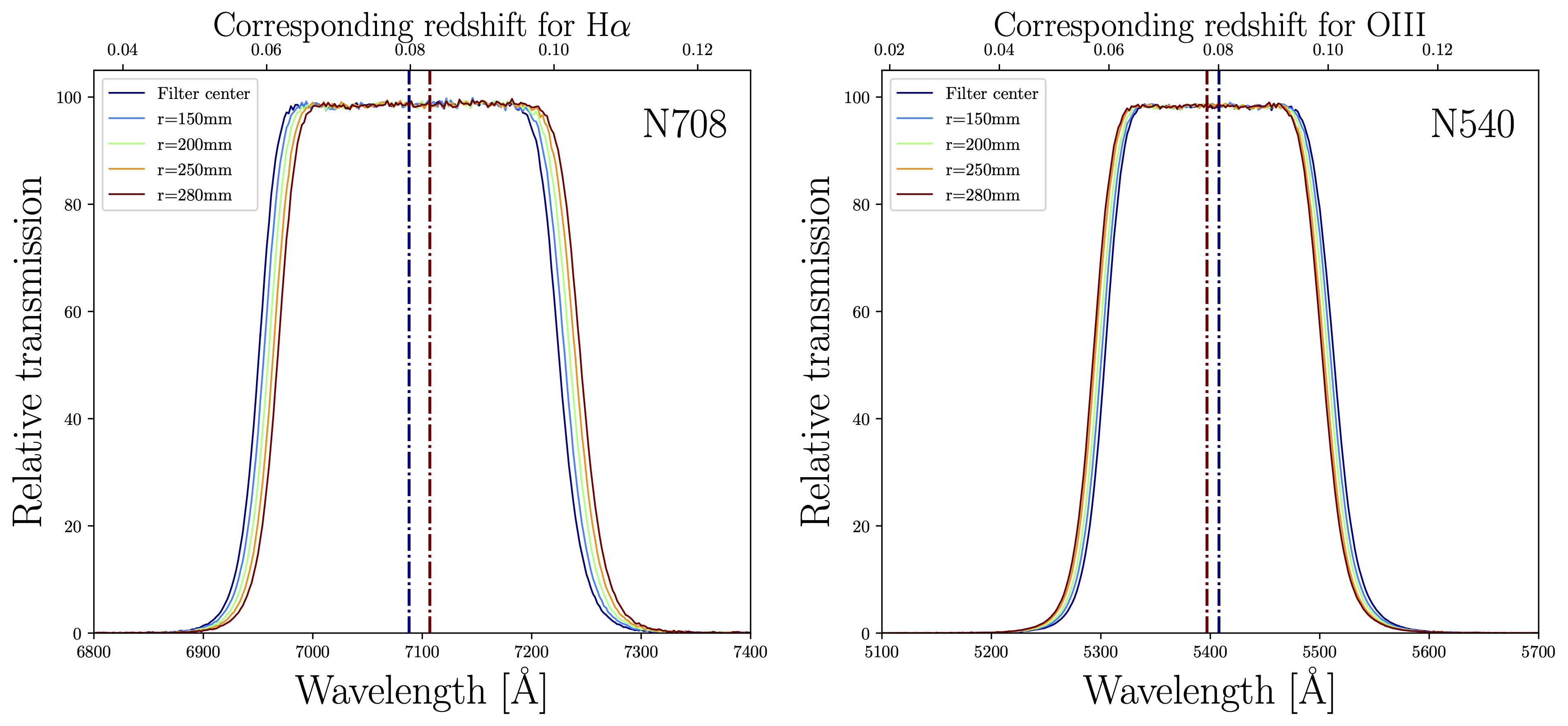}
\caption{Filter transmission curves for N708 and N540 measured in a parallel beam at AOI 5 degrees. There were measured by the filter
manufacturer Asahi Spectra Co. Blue, light blue, green, orange, and red solid curves are measured at positions that are 0 mm, 150 mm, 200 mm, 250 mm, and 280 mm away from the filter center. Blue and red dash-dotted lines show the central wavelength of the transmission curves at the filter center (r=0 mm) and filter edge (r=280 mm).}
\label{filter_asahi}
\end{center}
\end{figure*}

\subsection{Predicted photo-$z$ Performance}\label{merian_photoz}

We now discuss how much redshift information can be obtained with photo-$z$ fitting for galaxies in $0.058<z<0.10$. We characterize the expected photo-$z$ performance of the Merian 7-band design by generating mock samples of dwarfs and performing SED fitting on this mock sample. We focus on galaxies at $\rm z_{COSMOS}<0.2$ in this section since higher redshift interlopers can be removed by color-color cuts (e.g. $i-\rm N708>-0.1$ and $r-\rm N540>-0.2$), as shown in the previous section. This is a first indication of the Merian data, and the actual Merian selection will be studied and described in upcoming Merian papers.

We fit the SEDs of mock galaxies using {\tt Prospector} with the same model and priors described in Section~\ref{SED_lib}.  We run the SED fitting to derive photo-$z$'s, stellar masses, and star formation rates with 6-band ($grizy\rm N708$) and 7-band ($grizy\rm N708N540$) separately. Note that this is an optimistic scenario since we use the same galaxy Prospector galaxy model in the simulations and in the fitting process.

We now compare our results with HSC 5-band ($grizy$) photo-$z$'s. We use the HSC 5-band photo-$z$ based on the Bayesian template fitting-code {\tt Mizuki} for comparison \citep{Tanaka:2018}. Figure \ref{photoz_comparison} shows the comparison between HSC 5-band photo-$z$'s, our 7-band photo-$z$'s and $\rm z_{COSMOS}$, color-coded by HSC $g-i$ color as a star-formation rate indicator. The black boxes indicate our target redshift range of $0.058<z<0.1$. The 5-band HSC broad-band selection has a photo-$z$ accuracy of $\sigma_{\Delta z/(1+z)}\sim 0.5$ and an outlier fraction ($|z_{phot}-z_{spec}|/(1+z_{spec})>0.15$) of $\eta=60\%$. The 5-band selection below $z<0.1$ has very poor completeness and purity (48\% and 12\%), where completeness stands for how many dwarfs with $z_{\rm true}<0.1$ have $z_{\rm phot}<0.1$ and purity stands for how many dwarfs with $z_{\rm phot}<0.1$ are actually at $z_{\rm true}<0.1$.  The 7-band design yields $\sigma_{\Delta z/(1+z)}\sim 0.01$ and $\eta=2.8\%$ (this precision was also tested using galaxies in COSMOS with spectroscopic redshifts) and excellent completeness and purity (89\% and 90\%) in the target redshift range. The precision of the 6-band selection (not shown in the figure) is in between the 5-band and 7-band design, with $\sigma_{\Delta z/(1+z)}\sim 0.02$. 

The 7-band selection achieves a line-of-sight localization precision of $44.5$ Mpc (compared to 72.2 Mpc from the 6-band selection) within the $0.058<z<0.10$ window (Figure \ref{delta_d}). Most star-forming galaxies in our target redshift range (based on their $g-i$ color) have a 7-band photo-$z$ within the target redshift range $0.058<z<0.1$. We think that the 7-filter design improves redshifts over the 6-band design for two reasons: a) medium/narrow bands placed at the intersection between broad bands help to improve photo-$z$'s in general \citep[][]{Ilbert:2009}, and b) the N540 filter also provides redshift information as the $\rm H\alpha$, $\rm [OIII]$ and H$\beta$ lines move through the filters.

Our method is not optimized for redshift accuracy for quenched galaxies. However, the quenched fraction of dwarf galaxies in our target mass range is low. \citet{Geha:2012} found a quenched fraction for SDSS field dwarf galaxies with \merianmass\ of  less than  $0.06\%$. And recently, with the data from the Satellites Around Galactic Analogs Survey (SAGA), \citet{Mao:2021} found a quenched fraction for satellite dwarf galaxies around Milky Way analogs in this mass range (\merianmass) of $\sim10\%$. \citet{Carlsten:2022} report a higher satellite quenched fraction (15\% for \merianmass) using the Exploration of Local VolumE Satellites (ELVES) Survey. Since we are not selecting dwarf galaxies by their environments, we expect that the quenched fraction will be below $15\%$. In fact, 95\% of dwarf galaxies between \merianmass\ in our Initial Dwarf Sample sample are star-forming, according to the COSMOS2015 quiescent/star-forming classification, which is consistent with the quenched fraction found by \citet{Geha:2012} and \citet{Mao:2021}.  

The improved photo-$z$ precision of Merian over HSC 5 band photometry is important for selecting clean samples of dwarf galaxies and for measuring weak gravitational lensing for dwarfs. However, it is also important to note that this is an ideal simulation that does not account for photometric systematic errors (e.g. blending, galaxy morphology, photometric calibration, etc.). The same models were also used to generate the mock data and and to fit SEDs. The actual Merian photometric redshifts may therefore not achieve this theoretical performance.

\section{Construction and Characterization of the Merian Dual Filter System}\label{filter_construction}

\begin{figure*}
\begin{center}
\includegraphics[width=16cm]{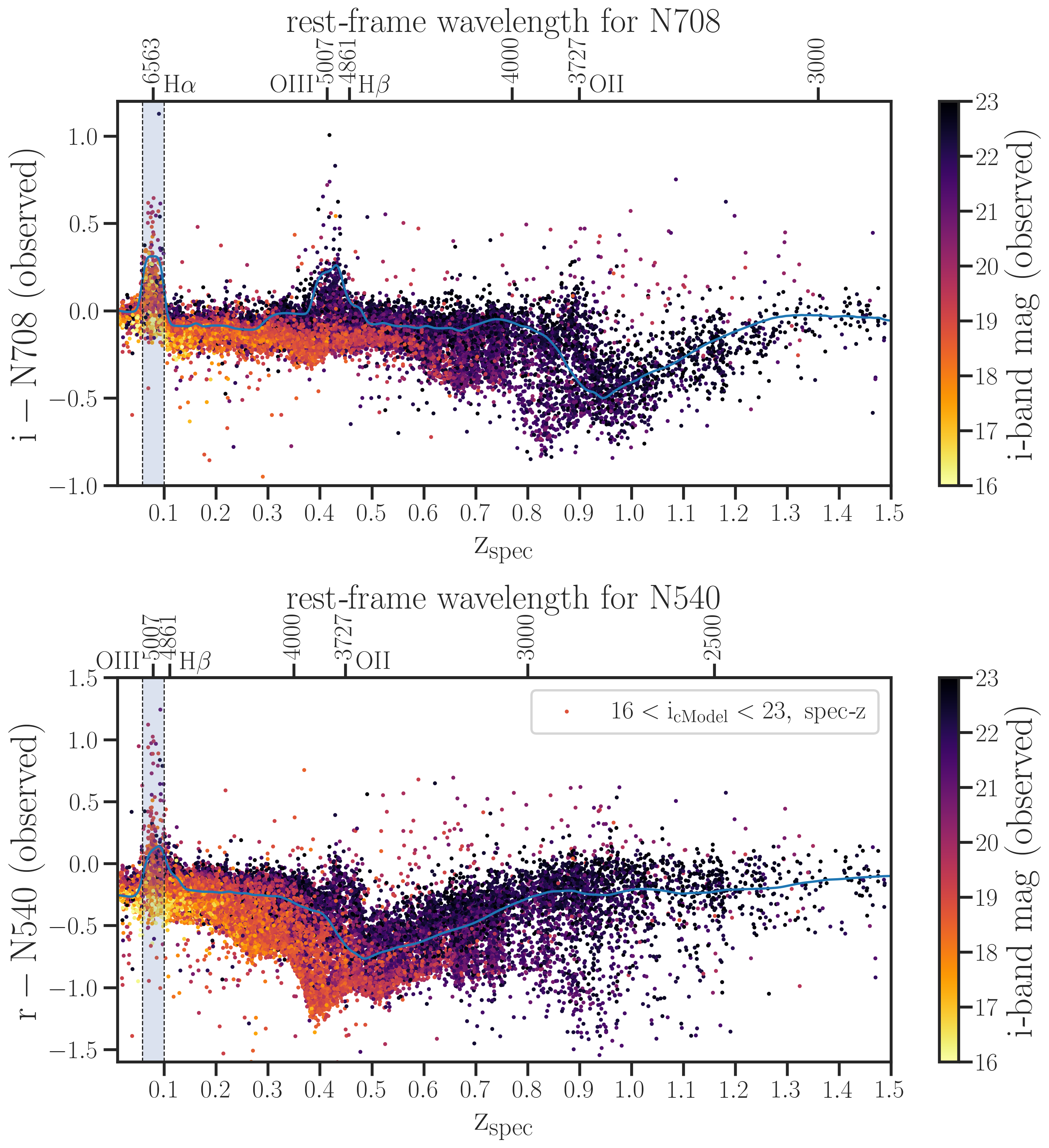}
\caption{Merian observed \iha\ and \roiii\ colors vs. redshift, color-coded by integrated cModel i-band magnitude for galaxies with $16<i<23$ and with a spectroscopic redshift. The blue solid curves show how \iha\ and \roiii\ change with redshift for a mock star-forming dwarf galaxy drawn from our dwarf SED library. The upper x-axis indicates the restframe wavelength probed by N708 and N540 (i.e. 7080\AA/(1+z) and 5400\AA/(1+z)) and indicates when  $\rm H\alpha 6563$ and $\rm [OIII]5007$ are in the bandpass. Shaded areas show the target Merian redshift range $0.058<z<0.1$.}
\label{flux_excess_observed}
\end{center}
\end{figure*}

Here we present the construction of the Merian filter set and show the actual performance of N708 and N540. We also present the first version of the Merian photometry, and highlight Merian's potential for finding dwarf galaxies at $0.058<z<0.1$ with actual data.

\subsection{Filter Construction and Performance}

The N708 and N540 filters were manufactured by Asahi Spectra Co. in 2020 and 2021 and then shipped to Chile and installed on DECam. Asahi Spectra Co. measured the filter transmission curves at various locations in the laboratory before the filters were shipped. These transmission curves are presented in Figure \ref{filter_asahi}. Color solid curves are filter transmission curves at  0 mm (i.e. center), 150 mm, 200 mm, 250 mm, and 280 mm (close to the edge and away from the filter center). The dash-dotted lines show the central wavelength of the transmission curves at the filter center ($r=0$ mm) and the filter edge ($r=280$ mm). The filter transmission curves shift by 19\AA\ for N708 and 11\AA\ for N540 from the filter center to the filter edge, which inevitably would also affect the corresponding redshift range. This effect will contribute to the systematic uncertainties of the photometry and will also add uncertainties to the final Merian photo-$z$.

\subsection{Merian First Year Data}\label{meriandata}

Merian observations have been ongoing since March 2021, and as such we can also provide the reader with a first look at the realized performance of the medium-band filters. We defer a full description of the observation and data reduction process to Danieli et al., \textit{in prep.}; here we will simply note that the broadband and medium-band data were reduced using the same pipeline, and that the $i-{\rm N708}$ and $r-{\rm N540}$ colors are corrected for differences in seeing between the HSC and DECam data.

In \autoref{flux_excess_observed} we show the real and observed equivalent of the mock color-redshift figure that we presented in \autoref{flux_excess_mock} for galaxies with spectroscpic redshifts and $16<i<23$. Here, redshifts are from a cross-match with the \textit{GAMA} spectroscopic survey and the COSMOS spectroscopic redshift catalog of Salvato et al., \textit{in prep}. We see good agreement between the predicted and observed distribution of medium-band to broadband colors as a function of spectroscopic redshift. This is demonstrated visually by the blue curve in \autoref{flux_excess_observed}, which traces the mean path of observed color evolution as  a function of redshift for a star-forming dwarf from our mock sample. As discussed above, the presence of other lines that move into the medium-bands at higher redshift (particularly $\rm [OIII]\lambda$5007\AA{} into N708 and the $\rm [OII]\lambda3727\AA,\lambda3729\AA$ into N540 at $z\approx0.4$) necessitates the use of a medium-band-assisted photometric redshift approach rather than a color cut.

Further analysis of the observed sample is beyond the scope of this work: here we simply underscore that a first look at the observed Merian galaxies shows that the behavior of our medium-band filters are indeed well-described by the mock catalog presented in this work.

\section{Summary and Conclusions}\label{conclusions}

Upcoming deep and wide imaging surveys open up the possibility of direct measurements of the dark matter halos of dwarf galaxies with weak gravitational lensing \citep[][]{Leauthaud:2020}. However, the key challenge towards achieving this goal is not the weak lensing measurement per se, but rather the selection of a reliable sample of dwarf galaxies from broad-band imaging data and the determination of accurate redshifts for these lenses. In this paper, we have optimized the design of two medium band filters (N708 and N540) to achieve this goal. The filter set designed in this paper forms the basis of the Merian survey which has been allocated 60 nights on the Blanco telescope to carry out an imaging survey of $\sim$ 850 degrees$^2$ and will perform the first high S/N measurements of gravitational lensing around dwarf galaxies.

The Merian filter design was optimized using a sample of dwarf galaxies from the COSMOS2015 catalog. We built a library of SEDs including emission lines for a complete sample of dwarf galaxies with \merianmass. We simulated the fluxes for these targets for various \Ha\ filter designs, varying both filter width $\Delta\lambda$ and filter central wavelength $\lambda_c$. Our design considers the predicted weak lensing S/N for dwarf galaxies as well as the magnitude difference between the Merian filters and HSC broad-band filters to measure photo-$z$'s. Our key design elements are summarized as follows.

\begin{itemize}

\item We first design our \Ha\ filter by optimizing the predicted weak lensing S/N. We force the detected fraction of dwarf galaxies to be more than 80$\%$, which sets the exposure time. The total number of dwarf galaxies used for lensing is then computed using a fixed number of nights (24 full nights for the \Ha\ filter). The predicted weak lensing S/N peaks at S/N$\sim$32 with a \Ha\ filter design $\Delta\lambda \sim$ 300\AA \ and $z_c\sim0.08$. For the filter central wavelength in \Ha\ redshift $z_c$, the predicted lensing S/N decreases above $z_c\sim 0.1$ because more distant, fainter galaxies need longer exposure times, resulting in a smaller survey area, cosmological volume, and lens sample. The lensing S/N also decreases below $z\sim 0.05$ because of the fixed area of the HSC-SSP wide layer ($\rm \sim 1000\ deg^{2}$). 

\item Merian will use photo-$z$/SED fitting to produce a reliable model of continuum in order to fit fluxes from emission lines, as well as to determine photo-$z$'s. The reliability of the model continuum from this fitting will depend on the photometric precision of Merian and HSC. We estimate the magnitude difference between Merian and HSC magnitudes as a function of filter width. Narrower filters have larger magnitude difference since they have less continuum to smear the emission line fluxes. We design a matched $\rm [OIII]$ filter with the same corresponding redshift range as the \Ha\ filter. We adjust the filter design to avoid a strong sky line at 5580\AA\ for the $\rm [OIII]$ filter. 

\item The optimization procedure led to the design of the Merian dual filter set: N708 ($\lambda_c = 7080$\AA, $\Delta\lambda = 275$\AA) and N540 ($\lambda_c = 5400$\AA, $\Delta\lambda = 210$\AA). The predicted lensing S/N for our final filter design is 32 at $r<0.5$ Mpc and 90 at $r<1.0$ Mpc, assuming a fixed observing time of 60 full nights (24 for \Ha\ and 36 for $\rm [OIII]$) on DECam. The exposure times for the \Ha\ and $\rm [OIII]$ filters are estimated by requiring the same depth according to our image simulations.

\item Using mock data we predict that Merian may be able to measure redshifts using 7-band photometry with a photometric redshift accuracy of $\sigma_{\Delta z/(1+z)}\sim 0.01$ and an outlier fraction of $\eta=2.8\%$ with excellent completeness and purity (89\% and 90\%) for dwarf galaxies with \merianmass\ at $0.058<z<0.10$. 

\end{itemize}

In this paper we also show early Merian photometry from our first year internal data release. We show that the emission lines are indeed detected in Merian filters with trends that mimic those predicted by our simulations. Science-ready photometry and photo-$z$ measurements are actively under development and will be presented in future work. 

The large sample of $\sim$ 85,000 star-forming dwarf galaxies that will be discovered by Merian will also allow for a wealth of ancillary studies of the properties of dwarf galaxies. This method of efficiently selecting low redshift dwarf galaxies from large imaging surveys can also be considered for other future imaging surveys, such as  {\it  Euclid} \citep{Laureijs:2011}, Nancy Grace Roman telescope \citep{Spergel:2013}, and the Legacy Survey of Space and Time (LSST) with the Rubin Observatory \citep{Ivezi:2019}. 

\section*{Acknowledgements}
This material is based upon work supported by the National Science Foundation under Grant No. 2106839.

We acknowledge use of the lux supercomputer at UC Santa Cruz, funded by NSF MRI grant AST 1828315. The authors are pleased to acknowledge that the work reported on in this paper was substantially performed using the Princeton Research Computing resources at Princeton University which is a consortium of groups led by the Princeton Institute for Computational Science and Engineering (PICSciE) and Office of Information Technology's Research Computing.

This research uses services or data provided by the Astro Data Lab at NSF’s National Optical-Infrared Astronomy
Research Laboratory. NOIRLab is operated by the Association of Universities for Research in Astronomy (AURA), Inc. under a cooperative agreement with the National Science Foundation.

This research makes use of software developed for the Vera C. Rubin Observatory. We thank the Vera C. Rubin Observatory for making their code available as free software at \url{http://dm.lsstcorp.org}.

The Hyper Suprime-Cam (HSC) collaboration includes the astronomical communities of Japan and Taiwan, and Princeton University. The HSC instrumentation and software were developed by National Astronomical Observatory of Japan (NAOJ), Kavli Institute for the Physics and Mathematics of the Universe (Kavli IPMU), University of Tokyo, High Energy Accelerator Research Organization (KEK), Academia Sinica Institute for Astronomy and Astrophysics in Taiwan (ASIAA), and Princeton University. Funding was contributed by the FIRST program from Japanese Cabinet Office, Ministry of Education, Culture, Sports, Science and Technology (MEXT), Japan Society for the Promotion of Science (JSPS), Japan Science and Technology Agency (JST), Toray Science Foundation, NAOJ, Kavli IPMU, KEK, ASIAA, and Princeton University.

This research made use of:
  \href{http://www.scipy.org/}{\texttt{SciPy}},
      an open source scientific tool for Python (\citealt{SciPy});
  \href{http://www.numpy.org/}{\texttt{NumPy}},
      a fundamental package for scientific computing with Python (\citealt{NumPy});
  \href{http://matplotlib.org/}{\texttt{Matplotlib}},
      a 2-D plotting library for Python (\citealt{Matplotlib});
  \href{http://www.astropy.org/}{\texttt{Astropy}}, a community-developed
      core Python package for astronomy (\citealt{astropy:2013, astropy:2018, astropy:2022});
  \href{http://galsim-developers.github.io/GalSim/_build/html/index.html}{\texttt{GalSim}},
      a modular galaxy image simulation toolkit in Python (\citealt{Rowe:2015}).
  \href{https://github.com/kbarbary/sep}{\texttt{sep}}
      Source Extraction and Photometry in Python (\citealt{SExtractor, PythonSEP});
  \href{https://dynesty.readthedocs.io/en/stable/}{\texttt{dynesty}},
      a dynamic nested sampling package for estimating Bayesian posteriors and evidences (\citealt{Speagle:2020});
  \href{https://www.star.bris.ac.uk/~mbt/topcat/}{\texttt{TOPCAT}},
      a user-friendly graphical program for viewing, analysis and editing of tables (\citealt{topcat}).

\section*{Data Availability}

The data underlying this article will be shared on reasonable request to the corresponding author.


\bibliographystyle{mnras}
\bibliography{dwarfmediumband}

\begin{thebibliography}{}
\makeatletter
\relax
\def\mn@urlcharsother{\let\do\@makeother \do\$\do\&\do\#\do\^\do\_\do\%\do\~}
\def\mn@doi{\begingroup\mn@urlcharsother \@ifnextchar [ {\mn@doi@}
  {\mn@doi@[]}}
\def\mn@doi@[#1]#2{\def\@tempa{#1}\ifx\@tempa\@empty \href
  {http://dx.doi.org/#2} {doi:#2}\else \href {http://dx.doi.org/#2} {#1}\fi
  \endgroup}
\def\mn@eprint#1#2{\mn@eprint@#1:#2::\@nil}
\def\mn@eprint@arXiv#1{\href {http://arxiv.org/abs/#1} {{\tt arXiv:#1}}}
\def\mn@eprint@dblp#1{\href {http://dblp.uni-trier.de/rec/bibtex/#1.xml}
  {dblp:#1}}
\def\mn@eprint@#1:#2:#3:#4\@nil{\def\@tempa {#1}\def\@tempb {#2}\def\@tempc
  {#3}\ifx \@tempc \@empty \let \@tempc \@tempb \let \@tempb \@tempa \fi \ifx
  \@tempb \@empty \def\@tempb {arXiv}\fi \@ifundefined
  {mn@eprint@\@tempb}{\@tempb:\@tempc}{\expandafter \expandafter \csname
  mn@eprint@\@tempb\endcsname \expandafter{\@tempc}}}

\bibitem[\protect\citeauthoryear{{Abbott} et~al.,}{{Abbott}
  et~al.}{2021}]{Abbott:2021}
{Abbott} T.~M.~C.,  et~al., 2021, \mn@doi [\apjs] {10.3847/1538-4365/ac00b3},
  \href {https://ui.adsabs.harvard.edu/abs/2021ApJS..255...20A} {255, 20}

\bibitem[\protect\citeauthoryear{{Aihara} et~al.,}{{Aihara}
  et~al.}{2018}]{Aihara:2018}
{Aihara} H.,  et~al., 2018, \mn@doi [\pasj] {10.1093/pasj/psx066}, \href
  {https://ui.adsabs.harvard.edu/abs/2018PASJ...70S...4A} {70, S4}

\bibitem[\protect\citeauthoryear{{Arnouts} et~al.,}{{Arnouts}
  et~al.}{2002}]{Arnouts:2002}
{Arnouts} S.,  et~al., 2002, \mn@doi [\mnras]
  {10.1046/j.1365-8711.2002.04988.x}, \href
  {https://ui.adsabs.harvard.edu/abs/2002MNRAS.329..355A} {329, 355}

\bibitem[\protect\citeauthoryear{{Astropy Collaboration} et~al.,}{{Astropy
  Collaboration} et~al.}{2013}]{astropy:2013}
{Astropy Collaboration} et~al., 2013, \mn@doi [\aap]
  {10.1051/0004-6361/201322068}, \href
  {http://adsabs.harvard.edu/abs/2013A%26A...558A..33A} {558, A33}

\bibitem[\protect\citeauthoryear{{Astropy Collaboration} et~al.,}{{Astropy
  Collaboration} et~al.}{2018}]{astropy:2018}
{Astropy Collaboration} et~al., 2018, \mn@doi [\aj] {10.3847/1538-3881/aabc4f},
  \href {https://ui.adsabs.harvard.edu/abs/2018AJ....156..123A} {156, 123}

\bibitem[\protect\citeauthoryear{{Astropy Collaboration} et~al.,}{{Astropy
  Collaboration} et~al.}{2022}]{astropy:2022}
{Astropy Collaboration} et~al., 2022, \mn@doi [apj] {10.3847/1538-4357/ac7c74},
  \href {https://ui.adsabs.harvard.edu/abs/2022ApJ...935..167A} {935, 167}

\bibitem[\protect\citeauthoryear{{Baldwin}, {Phillips}  \&
  {Terlevich}}{{Baldwin} et~al.}{1981}]{Baldwin:1981}
{Baldwin} J.~A.,  {Phillips} M.~M.,   {Terlevich} R.,  1981, \mn@doi [\pasp]
  {10.1086/130766}, \href
  {https://ui.adsabs.harvard.edu/abs/1981PASP...93....5B} {93, 5}

\bibitem[\protect\citeauthoryear{Barbary}{Barbary}{2016}]{PythonSEP}
Barbary K.,  2016, \mn@doi [Journal of Open Source Software]
  {10.21105/joss.00058}, 1, 58

\bibitem[\protect\citeauthoryear{Barbary, Boone  \& Deil}{Barbary
  et~al.}{2015}]{Barbary:2015}
Barbary K.,  Boone K.,   Deil C.,  2015, sep: v0.3.0,
  \mn@doi{10.5281/zenodo.15669}, \url {http://dx.doi.org/10.5281/zenodo.15669}

\bibitem[\protect\citeauthoryear{{Battaglia} \& {Nipoti}}{{Battaglia} \&
  {Nipoti}}{2022}]{Battaglia:2022}
{Battaglia} G.,  {Nipoti} C.,  2022, \mn@doi [Nature Astronomy]
  {10.1038/s41550-022-01638-7}, \href
  {https://ui.adsabs.harvard.edu/abs/2022NatAs...6..659B} {6, 659}

\bibitem[\protect\citeauthoryear{{Ben{\'\i}tez} et~al.,}{{Ben{\'\i}tez}
  et~al.}{2009}]{Benitez:2009}
{Ben{\'\i}tez} N.,  et~al., 2009, \mn@doi [\apj] {10.1088/0004-637X/691/1/241},
  \href {https://ui.adsabs.harvard.edu/abs/2009ApJ...691..241B} {691, 241}

\bibitem[\protect\citeauthoryear{{Benitez} et~al.,}{{Benitez}
  et~al.}{2014}]{Benitez:2014}
{Benitez} N.,  et~al., 2014, \mn@doi [arXiv e-prints]
  {10.48550/arXiv.1403.5237}, \href
  {https://ui.adsabs.harvard.edu/abs/2014arXiv1403.5237B} {p. arXiv:1403.5237}

\bibitem[\protect\citeauthoryear{{Bertin} \& {Arnouts}}{{Bertin} \&
  {Arnouts}}{1996}]{SExtractor}
{Bertin} E.,  {Arnouts} S.,  1996, \mn@doi [\aaps] {10.1051/aas:1996164}, \href
  {https://ui.adsabs.harvard.edu/abs/1996A&AS..117..393B} {117, 393}

\bibitem[\protect\citeauthoryear{{Bouch{\'e}} et~al.,}{{Bouch{\'e}}
  et~al.}{2022}]{Bouche:2022}
{Bouch{\'e}} N.~F.,  et~al., 2022, \mn@doi [\aap]
  {10.1051/0004-6361/202141762}, \href
  {https://ui.adsabs.harvard.edu/abs/2022A&A...658A..76B} {658, A76}

\bibitem[\protect\citeauthoryear{{Boylan-Kolchin}, {Bullock}  \&
  {Kaplinghat}}{{Boylan-Kolchin} et~al.}{2011}]{Boylan-Kolchin:2011}
{Boylan-Kolchin} M.,  {Bullock} J.~S.,   {Kaplinghat} M.,  2011, \mn@doi
  [\mnras] {10.1111/j.1745-3933.2011.01074.x}, \href
  {https://ui.adsabs.harvard.edu/abs/2011MNRAS.415L..40B} {415, L40}

\bibitem[\protect\citeauthoryear{{Boylan-Kolchin}, {Bullock}  \&
  {Kaplinghat}}{{Boylan-Kolchin} et~al.}{2012}]{Boylan-Kolchin:2012}
{Boylan-Kolchin} M.,  {Bullock} J.~S.,   {Kaplinghat} M.,  2012, \mn@doi
  [\mnras] {10.1111/j.1365-2966.2012.20695.x}, \href
  {https://ui.adsabs.harvard.edu/abs/2012MNRAS.422.1203B} {422, 1203}

\bibitem[\protect\citeauthoryear{{Brainerd}, {Blandford}  \&
  {Smail}}{{Brainerd} et~al.}{1996}]{Brainerd:1996}
{Brainerd} T.~G.,  {Blandford} R.~D.,   {Smail} I.,  1996, \mn@doi [\apj]
  {10.1086/177537}, \href
  {https://ui.adsabs.harvard.edu/abs/1996ApJ...466..623B} {466, 623}

\bibitem[\protect\citeauthoryear{{Brooks} \& {Zolotov}}{{Brooks} \&
  {Zolotov}}{2014}]{Brooks:2014}
{Brooks} A.~M.,  {Zolotov} A.,  2014, \mn@doi [\apj]
  {10.1088/0004-637X/786/2/87}, \href
  {https://ui.adsabs.harvard.edu/abs/2014ApJ...786...87B} {786, 87}

\bibitem[\protect\citeauthoryear{{Brooks}, {Papastergis}, {Christensen},
  {Governato}, {Stilp}, {Quinn}  \& {Wadsley}}{{Brooks}
  et~al.}{2017}]{Brooks:2017}
{Brooks} A.~M.,  {Papastergis} E.,  {Christensen} C.~R.,  {Governato} F.,
  {Stilp} A.,  {Quinn} T.~R.,   {Wadsley} J.,  2017, \mn@doi [\apj]
  {10.3847/1538-4357/aa9576}, \href
  {https://ui.adsabs.harvard.edu/abs/2017ApJ...850...97B} {850, 97}

\bibitem[\protect\citeauthoryear{{Buckley} \& {Peter}}{{Buckley} \&
  {Peter}}{2018}]{Buckley:2018}
{Buckley} M.~R.,  {Peter} A. H.~G.,  2018, \mn@doi [\physrep]
  {10.1016/j.physrep.2018.07.003}, \href
  {https://ui.adsabs.harvard.edu/abs/2018PhR...761....1B} {761, 1}

\bibitem[\protect\citeauthoryear{{Bullock} \& {Boylan-Kolchin}}{{Bullock} \&
  {Boylan-Kolchin}}{2017}]{Bullock:2017}
{Bullock} J.~S.,  {Boylan-Kolchin} M.,  2017, \mn@doi [\araa]
  {10.1146/annurev-astro-091916-055313}, \href
  {https://ui.adsabs.harvard.edu/abs/2017ARA&A..55..343B} {55, 343}

\bibitem[\protect\citeauthoryear{{Byler}, {Dalcanton}, {Conroy}  \&
  {Johnson}}{{Byler} et~al.}{2017}]{Byler:2017}
{Byler} N.,  {Dalcanton} J.~J.,  {Conroy} C.,   {Johnson} B.~D.,  2017, \mn@doi
  [\apj] {10.3847/1538-4357/aa6c66}, \href
  {https://ui.adsabs.harvard.edu/abs/2017ApJ...840...44B} {840, 44}

\bibitem[\protect\citeauthoryear{{Calzetti}, {Armus}, {Bohlin}, {Kinney},
  {Koornneef}  \& {Storchi-Bergmann}}{{Calzetti} et~al.}{2000}]{Calzetti:2000}
{Calzetti} D.,  {Armus} L.,  {Bohlin} R.~C.,  {Kinney} A.~L.,  {Koornneef} J.,
   {Storchi-Bergmann} T.,  2000, \mn@doi [\apj] {10.1086/308692}, \href
  {https://ui.adsabs.harvard.edu/abs/2000ApJ...533..682C} {533, 682}

\bibitem[\protect\citeauthoryear{{Carlsten}, {Greene}, {Beaton}, {Danieli}  \&
  {Greco}}{{Carlsten} et~al.}{2022}]{Carlsten:2022}
{Carlsten} S.~G.,  {Greene} J.~E.,  {Beaton} R.~L.,  {Danieli} S.,   {Greco}
  J.~P.,  2022, arXiv e-prints, \href
  {https://ui.adsabs.harvard.edu/abs/2022arXiv220300014C} {p. arXiv:2203.00014}

\bibitem[\protect\citeauthoryear{{Chabrier}}{{Chabrier}}{2003}]{Chabrier:2003}
{Chabrier} G.,  2003, \mn@doi [\pasp] {10.1086/376392}, \href
  {https://ui.adsabs.harvard.edu/abs/2003PASP..115..763C} {115, 763}

\bibitem[\protect\citeauthoryear{{Choi}, {Dotter}, {Conroy}, {Cantiello},
  {Paxton}  \& {Johnson}}{{Choi} et~al.}{2016}]{Choi:2016}
{Choi} J.,  {Dotter} A.,  {Conroy} C.,  {Cantiello} M.,  {Paxton} B.,
  {Johnson} B.~D.,  2016, \mn@doi [\apj] {10.3847/0004-637X/823/2/102}, \href
  {https://ui.adsabs.harvard.edu/abs/2016ApJ...823..102C} {823, 102}

\bibitem[\protect\citeauthoryear{{Civano} et~al.,}{{Civano}
  et~al.}{2016}]{Civano:2016}
{Civano} F.,  et~al., 2016, \mn@doi [\apj] {10.3847/0004-637X/819/1/62}, \href
  {https://ui.adsabs.harvard.edu/abs/2016ApJ...819...62C} {819, 62}

\bibitem[\protect\citeauthoryear{{Coil} et~al.,}{{Coil}
  et~al.}{2011}]{Coil:2011}
{Coil} A.~L.,  et~al., 2011, \mn@doi [\apj] {10.1088/0004-637X/741/1/8}, \href
  {https://ui.adsabs.harvard.edu/abs/2011ApJ...741....8C} {741, 8}

\bibitem[\protect\citeauthoryear{{Collins} \& {Read}}{{Collins} \&
  {Read}}{2022}]{Collins:2022}
{Collins} M. L.~M.,  {Read} J.~I.,  2022, \mn@doi [Nature Astronomy]
  {10.1038/s41550-022-01657-4}, \href
  {https://ui.adsabs.harvard.edu/abs/2022NatAs...6..647C} {6, 647}

\bibitem[\protect\citeauthoryear{{Conroy} \& {Gunn}}{{Conroy} \&
  {Gunn}}{2010}]{Conroy:2010}
{Conroy} C.,  {Gunn} J.~E.,  2010, \mn@doi [\apj]
  {10.1088/0004-637X/712/2/833}, \href
  {https://ui.adsabs.harvard.edu/abs/2010ApJ...712..833C} {712, 833}

\bibitem[\protect\citeauthoryear{{Conroy}, {Gunn}  \& {White}}{{Conroy}
  et~al.}{2009}]{Conroy:2009}
{Conroy} C.,  {Gunn} J.~E.,   {White} M.,  2009, \mn@doi [\apj]
  {10.1088/0004-637X/699/1/486}, \href
  {https://ui.adsabs.harvard.edu/abs/2009ApJ...699..486C} {699, 486}

\bibitem[\protect\citeauthoryear{{Coupon}, {Czakon}, {Bosch}, {Komiyama},
  {Medezinski}, {Miyazaki}  \& {Oguri}}{{Coupon} et~al.}{2018}]{Coupon:2018}
{Coupon} J.,  {Czakon} N.,  {Bosch} J.,  {Komiyama} Y.,  {Medezinski} E.,
  {Miyazaki} S.,   {Oguri} M.,  2018, \mn@doi [\pasj] {10.1093/pasj/psx047},
  \href {https://ui.adsabs.harvard.edu/abs/2018PASJ...70S...7C} {70, S7}

\bibitem[\protect\citeauthoryear{{Danieli}, {Greene}, {Carlsten}, {Jiang},
  {Beaton}  \& {Goulding}}{{Danieli} et~al.}{2022}]{Danieli:2022}
{Danieli} S.,  {Greene} J.~E.,  {Carlsten} S.,  {Jiang} F.,  {Beaton} R.,
  {Goulding} A.~D.,  2022, \mn@doi [arXiv e-prints]
  {10.48550/arXiv.2210.14233}, \href
  {https://ui.adsabs.harvard.edu/abs/2022arXiv221014233D} {p. arXiv:2210.14233}

\bibitem[\protect\citeauthoryear{{Darragh-Ford} et~al.,}{{Darragh-Ford}
  et~al.}{2022}]{Darragh-Ford:2022}
{Darragh-Ford} E.,  et~al., 2022, \mn@doi [arXiv e-prints]
  {10.48550/arXiv.2212.07433}, \href
  {https://ui.adsabs.harvard.edu/abs/2022arXiv221207433D} {p. arXiv:2212.07433}

\bibitem[\protect\citeauthoryear{{Davidzon} et~al.,}{{Davidzon}
  et~al.}{2017}]{Davidzon:2017}
{Davidzon} I.,  et~al., 2017, \mn@doi [\aap] {10.1051/0004-6361/201730419},
  \href {https://ui.adsabs.harvard.edu/abs/2017A&A...605A..70D} {605, A70}

\bibitem[\protect\citeauthoryear{{De Leo}, {Read}, {Noel}, {Erkal}, {Massana}
  \& {Carrera}}{{De Leo} et~al.}{2023}]{DeLeo:2023}
{De Leo} M.,  {Read} J.~I.,  {Noel} N. E.~D.,  {Erkal} D.,  {Massana} P.,
  {Carrera} R.,  2023, \mn@doi [arXiv e-prints] {10.48550/arXiv.2303.08838},
  \href {https://ui.adsabs.harvard.edu/abs/2023arXiv230308838D} {p.
  arXiv:2303.08838}

\bibitem[\protect\citeauthoryear{{Dey} et~al.,}{{Dey} et~al.}{2019}]{Dey:2019}
{Dey} A.,  et~al., 2019, \mn@doi [\aj] {10.3847/1538-3881/ab089d}, \href
  {https://ui.adsabs.harvard.edu/abs/2019AJ....157..168D} {157, 168}

\bibitem[\protect\citeauthoryear{{Di Cintio}, {Brook}, {Macci{\`o}}, {Stinson},
  {Knebe}, {Dutton}  \& {Wadsley}}{{Di Cintio} et~al.}{2014a}]{DiCintio:2014a}
{Di Cintio} A.,  {Brook} C.~B.,  {Macci{\`o}} A.~V.,  {Stinson} G.~S.,  {Knebe}
  A.,  {Dutton} A.~A.,   {Wadsley} J.,  2014a, \mn@doi [\mnras]
  {10.1093/mnras/stt1891}, \href
  {https://ui.adsabs.harvard.edu/abs/2014MNRAS.437..415D} {437, 415}

\bibitem[\protect\citeauthoryear{{Di Cintio}, {Brook}, {Dutton}, {Macci{\`o}},
  {Stinson}  \& {Knebe}}{{Di Cintio} et~al.}{2014b}]{DiCintio:2014b}
{Di Cintio} A.,  {Brook} C.~B.,  {Dutton} A.~A.,  {Macci{\`o}} A.~V.,
  {Stinson} G.~S.,   {Knebe} A.,  2014b, \mn@doi [\mnras]
  {10.1093/mnras/stu729}, \href
  {https://ui.adsabs.harvard.edu/abs/2014MNRAS.441.2986D} {441, 2986}

\bibitem[\protect\citeauthoryear{{Drlica-Wagner} et~al.,}{{Drlica-Wagner}
  et~al.}{2022}]{Drlica-Wagner:2022}
{Drlica-Wagner} A.,  et~al., 2022, \mn@doi [\apjs] {10.3847/1538-4365/ac78eb},
  \href {https://ui.adsabs.harvard.edu/abs/2022ApJS..261...38D} {261, 38}

\bibitem[\protect\citeauthoryear{{Dutton}}{{Dutton}}{2009}]{Dutton:2009}
{Dutton} A.~A.,  2009, \mn@doi [\mnras] {10.1111/j.1365-2966.2009.14741.x},
  \href {https://ui.adsabs.harvard.edu/abs/2009MNRAS.396..121D} {396, 121}

\bibitem[\protect\citeauthoryear{{Dvornik} et~al.,}{{Dvornik}
  et~al.}{2022}]{Dvornik:2022}
{Dvornik} A.,  et~al., 2022, \mn@doi [arXiv e-prints]
  {10.48550/arXiv.2210.03110}, \href
  {https://ui.adsabs.harvard.edu/abs/2022arXiv221003110D} {p. arXiv:2210.03110}

\bibitem[\protect\citeauthoryear{{Eigenthaler} et~al.,}{{Eigenthaler}
  et~al.}{2018}]{Eigenthaler:2018}
{Eigenthaler} P.,  et~al., 2018, \mn@doi [\apj] {10.3847/1538-4357/aaab60},
  \href {https://ui.adsabs.harvard.edu/abs/2018ApJ...855..142E} {855, 142}

\bibitem[\protect\citeauthoryear{{Fisher} \& {Drory}}{{Fisher} \&
  {Drory}}{2011}]{Fisher:2011}
{Fisher} D.~B.,  {Drory} N.,  2011, \mn@doi [\apjl]
  {10.1088/2041-8205/733/2/L47}, \href
  {https://ui.adsabs.harvard.edu/abs/2011ApJ...733L..47F} {733, L47}

\bibitem[\protect\citeauthoryear{{Fitts} et~al.,}{{Fitts}
  et~al.}{2019}]{Fitts:2019}
{Fitts} A.,  et~al., 2019, \mn@doi [\mnras] {10.1093/mnras/stz2613}, \href
  {https://ui.adsabs.harvard.edu/abs/2019MNRAS.490..962F} {490, 962}

\bibitem[\protect\citeauthoryear{{Flores} \& {Primack}}{{Flores} \&
  {Primack}}{1994}]{Flores:1994}
{Flores} R.~A.,  {Primack} J.~R.,  1994, \mn@doi [\apjl] {10.1086/187350},
  \href {https://ui.adsabs.harvard.edu/abs/1994ApJ...427L...1F} {427, L1}

\bibitem[\protect\citeauthoryear{{Fry} et~al.,}{{Fry} et~al.}{2015}]{Fry:2015}
{Fry} A.~B.,  et~al., 2015, \mn@doi [\mnras] {10.1093/mnras/stv1330}, \href
  {https://ui.adsabs.harvard.edu/abs/2015MNRAS.452.1468F} {452, 1468}

\bibitem[\protect\citeauthoryear{{Geha}, {Blanton}, {Yan}  \& {Tinker}}{{Geha}
  et~al.}{2012}]{Geha:2012}
{Geha} M.,  {Blanton} M.~R.,  {Yan} R.,   {Tinker} J.~L.,  2012, \mn@doi [\apj]
  {10.1088/0004-637X/757/1/85}, \href
  {https://ui.adsabs.harvard.edu/abs/2012ApJ...757...85G} {757, 85}

\bibitem[\protect\citeauthoryear{{Governato} et~al.,}{{Governato}
  et~al.}{2010}]{Governato:2010}
{Governato} F.,  et~al., 2010, \mn@doi [\nat] {10.1038/nature08640}, \href
  {https://ui.adsabs.harvard.edu/abs/2010Natur.463..203G} {463, 203}

\bibitem[\protect\citeauthoryear{{Governato} et~al.,}{{Governato}
  et~al.}{2012}]{Governato:2012}
{Governato} F.,  et~al., 2012, \mn@doi [\mnras]
  {10.1111/j.1365-2966.2012.20696.x}, \href
  {https://ui.adsabs.harvard.edu/abs/2012MNRAS.422.1231G} {422, 1231}

\bibitem[\protect\citeauthoryear{Harris et~al.,}{Harris et~al.}{2020}]{NumPy}
Harris C.~R.,  et~al., 2020, \mn@doi [Nature] {10.1038/s41586-020-2649-2}, 585,
  357

\bibitem[\protect\citeauthoryear{{Hasinger} et~al.,}{{Hasinger}
  et~al.}{2018}]{Hasinger:2018}
{Hasinger} G.,  et~al., 2018, \mn@doi [\apj] {10.3847/1538-4357/aabacf}, \href
  {https://ui.adsabs.harvard.edu/abs/2018ApJ...858...77H} {858, 77}

\bibitem[\protect\citeauthoryear{{Heymans} et~al.,}{{Heymans}
  et~al.}{2012}]{Heymans:2012}
{Heymans} C.,  et~al., 2012, \mn@doi [\mnras]
  {10.1111/j.1365-2966.2012.21952.x}, \href
  {https://ui.adsabs.harvard.edu/abs/2012MNRAS.427..146H} {427, 146}

\bibitem[\protect\citeauthoryear{{Hikage} et~al.,}{{Hikage}
  et~al.}{2019}]{Hikage:2019}
{Hikage} C.,  et~al., 2019, \mn@doi [\pasj] {10.1093/pasj/psz010}, \href
  {https://ui.adsabs.harvard.edu/abs/2019PASJ...71...43H} {71, 43}

\bibitem[\protect\citeauthoryear{{Huang} et~al.,}{{Huang}
  et~al.}{2018}]{Huang:2018}
{Huang} S.,  et~al., 2018, \mn@doi [\pasj] {10.1093/pasj/psx126}, \href
  {https://ui.adsabs.harvard.edu/abs/2018PASJ...70S...6H} {70, S6}

\bibitem[\protect\citeauthoryear{{Hudson} et~al.,}{{Hudson}
  et~al.}{2015}]{Hudson:2015}
{Hudson} M.~J.,  et~al., 2015, \mn@doi [\mnras] {10.1093/mnras/stu2367}, \href
  {https://ui.adsabs.harvard.edu/abs/2015MNRAS.447..298H} {447, 298}

\bibitem[\protect\citeauthoryear{Hunter}{Hunter}{2007}]{Matplotlib}
Hunter J.~D.,  2007, \mn@doi [Computing in Science \& Engineering]
  {10.1109/MCSE.2007.55}, 9, 90

\bibitem[\protect\citeauthoryear{{Ilbert} et~al.,}{{Ilbert}
  et~al.}{2006}]{Ilbert:2006}
{Ilbert} O.,  et~al., 2006, \mn@doi [\aap] {10.1051/0004-6361:20065138}, \href
  {https://ui.adsabs.harvard.edu/abs/2006A&A...457..841I} {457, 841}

\bibitem[\protect\citeauthoryear{{Ilbert} et~al.,}{{Ilbert}
  et~al.}{2009}]{Ilbert:2009}
{Ilbert} O.,  et~al., 2009, \mn@doi [\apj] {10.1088/0004-637X/690/2/1236},
  \href {https://ui.adsabs.harvard.edu/abs/2009ApJ...690.1236I} {690, 1236}

\bibitem[\protect\citeauthoryear{{Ivezi{\'c}} et~al.,}{{Ivezi{\'c}}
  et~al.}{2019}]{Ivezi:2019}
{Ivezi{\'c}} {\v{Z}}.,  et~al., 2019, \mn@doi [\apj]
  {10.3847/1538-4357/ab042c}, \href
  {https://ui.adsabs.harvard.edu/abs/2019ApJ...873..111I} {873, 111}

\bibitem[\protect\citeauthoryear{{Johnson}, {Leja}, {Conroy}  \&
  {Speagle}}{{Johnson} et~al.}{2021}]{Johnson:2021}
{Johnson} B.~D.,  {Leja} J.,  {Conroy} C.,   {Speagle} J.~S.,  2021, \mn@doi
  [\apjs] {10.3847/1538-4365/abef67}, \href
  {https://ui.adsabs.harvard.edu/abs/2021ApJS..254...22J} {254, 22}

\bibitem[\protect\citeauthoryear{{Kado-Fong}, {Greene}, {Huang}, {Beaton},
  {Goulding}  \& {Komiyama}}{{Kado-Fong} et~al.}{2020}]{KadoFong:2020}
{Kado-Fong} E.,  {Greene} J.~E.,  {Huang} S.,  {Beaton} R.,  {Goulding} A.~D.,
   {Komiyama} Y.,  2020, \mn@doi [\apj] {10.3847/1538-4357/abacc2}, \href
  {https://ui.adsabs.harvard.edu/abs/2020ApJ...900..163K} {900, 163}

\bibitem[\protect\citeauthoryear{{Kado-Fong} et~al.,}{{Kado-Fong}
  et~al.}{2021}]{KadoFong:2021}
{Kado-Fong} E.,  et~al., 2021, \mn@doi [\apj] {10.3847/1538-4357/ac15f0}, \href
  {https://ui.adsabs.harvard.edu/abs/2021ApJ...920...72K} {920, 72}

\bibitem[\protect\citeauthoryear{{Kamada}, {Kaplinghat}, {Pace}  \&
  {Yu}}{{Kamada} et~al.}{2017}]{Kamada:2017}
{Kamada} A.,  {Kaplinghat} M.,  {Pace} A.~B.,   {Yu} H.-B.,  2017, \mn@doi
  [\prl] {10.1103/PhysRevLett.119.111102}, \href
  {https://ui.adsabs.harvard.edu/abs/2017PhRvL.119k1102K} {119, 111102}

\bibitem[\protect\citeauthoryear{{Kashino} et~al.,}{{Kashino}
  et~al.}{2019}]{Kashino:2019}
{Kashino} D.,  et~al., 2019, \mn@doi [\apjs] {10.3847/1538-4365/ab06c4}, \href
  {https://ui.adsabs.harvard.edu/abs/2019ApJS..241...10K} {241, 10}

\bibitem[\protect\citeauthoryear{{Klypin}, {Kravtsov}, {Valenzuela}  \&
  {Prada}}{{Klypin} et~al.}{1999}]{Klypin:1999}
{Klypin} A.,  {Kravtsov} A.~V.,  {Valenzuela} O.,   {Prada} F.,  1999, \mn@doi
  [\apj] {10.1086/307643}, \href
  {https://ui.adsabs.harvard.edu/abs/1999ApJ...522...82K} {522, 82}

\bibitem[\protect\citeauthoryear{{Klypin}, {Karachentsev}, {Makarov}  \&
  {Nasonova}}{{Klypin} et~al.}{2015}]{Klypin:2015}
{Klypin} A.,  {Karachentsev} I.,  {Makarov} D.,   {Nasonova} O.,  2015, \mn@doi
  [\mnras] {10.1093/mnras/stv2040}, \href
  {https://ui.adsabs.harvard.edu/abs/2015MNRAS.454.1798K} {454, 1798}

\bibitem[\protect\citeauthoryear{{Kravtsov}}{{Kravtsov}}{2013}]{Kravtsov:2013}
{Kravtsov} A.~V.,  2013, \mn@doi [\apjl] {10.1088/2041-8205/764/2/L31}, \href
  {https://ui.adsabs.harvard.edu/abs/2013ApJ...764L..31K} {764, L31}

\bibitem[\protect\citeauthoryear{{Laigle} et~al.,}{{Laigle}
  et~al.}{2016}]{Laigle:2016}
{Laigle} C.,  et~al., 2016, \mn@doi [\apjs] {10.3847/0067-0049/224/2/24}, \href
  {https://ui.adsabs.harvard.edu/abs/2016ApJS..224...24L} {224, 24}

\bibitem[\protect\citeauthoryear{{Laureijs} et~al.,}{{Laureijs}
  et~al.}{2011}]{Laureijs:2011}
{Laureijs} R.,  et~al., 2011, arXiv e-prints, \href
  {https://ui.adsabs.harvard.edu/abs/2011arXiv1110.3193L} {p. arXiv:1110.3193}

\bibitem[\protect\citeauthoryear{{Le F{\`e}vre} et~al.,}{{Le F{\`e}vre}
  et~al.}{2013}]{LeFevre2013}
{Le F{\`e}vre} O.,  et~al., 2013, \mn@doi [\aap] {10.1051/0004-6361/201322179},
  \href {https://ui.adsabs.harvard.edu/abs/2013A&A...559A..14L} {559, A14}

\bibitem[\protect\citeauthoryear{{Le F{\`e}vre} et~al.,}{{Le F{\`e}vre}
  et~al.}{2015}]{LeFevre:2015}
{Le F{\`e}vre} O.,  et~al., 2015, \mn@doi [\aap] {10.1051/0004-6361/201423829},
  \href {https://ui.adsabs.harvard.edu/abs/2015A&A...576A..79L} {576, A79}

\bibitem[\protect\citeauthoryear{{Le Floc'h} et~al.,}{{Le Floc'h}
  et~al.}{2009}]{LeFloch:2009}
{Le Floc'h} E.,  et~al., 2009, \mn@doi [\apj] {10.1088/0004-637X/703/1/222},
  \href {https://ui.adsabs.harvard.edu/abs/2009ApJ...703..222L} {703, 222}

\bibitem[\protect\citeauthoryear{{Leauthaud} et~al.,}{{Leauthaud}
  et~al.}{2007}]{Leauthaud:2007}
{Leauthaud} A.,  et~al., 2007, \mn@doi [\apjs] {10.1086/516598}, \href
  {https://ui.adsabs.harvard.edu/abs/2007ApJS..172..219L} {172, 219}

\bibitem[\protect\citeauthoryear{{Leauthaud} et~al.,}{{Leauthaud}
  et~al.}{2012}]{Leauthaud:2012}
{Leauthaud} A.,  et~al., 2012, \mn@doi [\apj] {10.1088/0004-637X/744/2/159},
  \href {http://adsabs.harvard.edu/abs/2012ApJ...744..159L} {744, 159}

\bibitem[\protect\citeauthoryear{{Leauthaud}, {Singh}, {Luo}, {Ardila},
  {Greco}, {Capak}, {Greene}  \& {Mayer}}{{Leauthaud}
  et~al.}{2020}]{Leauthaud:2020}
{Leauthaud} A.,  {Singh} S.,  {Luo} Y.,  {Ardila} F.,  {Greco} J.~P.,  {Capak}
  P.,  {Greene} J.~E.,   {Mayer} L.,  2020, \mn@doi [Physics of the Dark
  Universe] {10.1016/j.dark.2020.100719}, \href
  {https://ui.adsabs.harvard.edu/abs/2020PDU....3000719L} {30, 100719}

\bibitem[\protect\citeauthoryear{{Lee} et~al.,}{{Lee} et~al.}{2024}]{Lee:2024}
{Lee} K.-S.,  et~al., 2024, \mn@doi [\apj] {10.3847/1538-4357/ad165e}, \href
  {https://ui.adsabs.harvard.edu/abs/2024ApJ...962...36L} {962, 36}

\bibitem[\protect\citeauthoryear{{Lelli}}{{Lelli}}{2022}]{Lelli:2022}
{Lelli} F.,  2022, \mn@doi [Nature Astronomy] {10.1038/s41550-021-01562-2},
  \href {https://ui.adsabs.harvard.edu/abs/2022NatAs...6...35L} {6, 35}

\bibitem[\protect\citeauthoryear{{Lilly} et~al.,}{{Lilly}
  et~al.}{2007}]{Lilly:2007}
{Lilly} S.~J.,  et~al., 2007, \mn@doi [\apjs] {10.1086/516589}, \href
  {https://ui.adsabs.harvard.edu/abs/2007ApJS..172...70L} {172, 70}

\bibitem[\protect\citeauthoryear{{Liske} et~al.,}{{Liske}
  et~al.}{2015}]{Liske:2015}
{Liske} J.,  et~al., 2015, \mn@doi [\mnras] {10.1093/mnras/stv1436}, \href
  {https://ui.adsabs.harvard.edu/abs/2015MNRAS.452.2087L} {452, 2087}

\bibitem[\protect\citeauthoryear{{Lutz} et~al.,}{{Lutz}
  et~al.}{2011}]{Lutz:2011}
{Lutz} D.,  et~al., 2011, \mn@doi [\aap] {10.1051/0004-6361/201117107}, \href
  {https://ui.adsabs.harvard.edu/abs/2011A&A...532A..90L} {532, A90}

\bibitem[\protect\citeauthoryear{{Mao}, {Geha}, {Wechsler}, {Weiner},
  {Tollerud}, {Nadler}  \& {Kallivayalil}}{{Mao} et~al.}{2021}]{Mao:2021}
{Mao} Y.-Y.,  {Geha} M.,  {Wechsler} R.~H.,  {Weiner} B.,  {Tollerud} E.~J.,
  {Nadler} E.~O.,   {Kallivayalil} N.,  2021, \mn@doi [\apj]
  {10.3847/1538-4357/abce58}, \href
  {https://ui.adsabs.harvard.edu/abs/2021ApJ...907...85M} {907, 85}

\bibitem[\protect\citeauthoryear{{Mashchenko}, {Wadsley}  \&
  {Couchman}}{{Mashchenko} et~al.}{2008}]{Mashchenko:2008}
{Mashchenko} S.,  {Wadsley} J.,   {Couchman} H.~M.~P.,  2008, \mn@doi [Science]
  {10.1126/science.1148666}, \href
  {https://ui.adsabs.harvard.edu/abs/2008Sci...319..174M} {319, 174}

\bibitem[\protect\citeauthoryear{{Masters} et~al.,}{{Masters}
  et~al.}{2019}]{Masters:2019}
{Masters} D.~C.,  et~al., 2019, \mn@doi [\apj] {10.3847/1538-4357/ab184d},
  \href {https://ui.adsabs.harvard.edu/abs/2019ApJ...877...81M} {877, 81}

\bibitem[\protect\citeauthoryear{{McCracken} et~al.,}{{McCracken}
  et~al.}{2012}]{McCracken:2012}
{McCracken} H.~J.,  et~al., 2012, \mn@doi [\aap] {10.1051/0004-6361/201219507},
  \href {https://ui.adsabs.harvard.edu/abs/2012A&A...544A.156M} {544, A156}

\bibitem[\protect\citeauthoryear{{McQuinn} et~al.,}{{McQuinn}
  et~al.}{2022}]{McQuinn:2022}
{McQuinn} K. B.~W.,  et~al., 2022, arXiv e-prints, \href
  {https://ui.adsabs.harvard.edu/abs/2022arXiv220310105M} {p. arXiv:2203.10105}

\bibitem[\protect\citeauthoryear{{Mendes de Oliveira} et~al.,}{{Mendes de
  Oliveira} et~al.}{2019}]{deOliveira:2019}
{Mendes de Oliveira} C.,  et~al., 2019, \mn@doi [\mnras]
  {10.1093/mnras/stz1985}, \href
  {https://ui.adsabs.harvard.edu/abs/2019MNRAS.489..241M} {489, 241}

\bibitem[\protect\citeauthoryear{{Miller}, {Ellis}, {Newman}  \&
  {Benson}}{{Miller} et~al.}{2014}]{Miller:2014}
{Miller} S.~H.,  {Ellis} R.~S.,  {Newman} A.~B.,   {Benson} A.,  2014, \mn@doi
  [\apj] {10.1088/0004-637X/782/2/115}, \href
  {https://ui.adsabs.harvard.edu/abs/2014ApJ...782..115M} {782, 115}

\bibitem[\protect\citeauthoryear{{Miralda-Escude}}{{Miralda-Escude}}{1991}]{Miralda-Escude:1991}
{Miralda-Escude} J.,  1991, \mn@doi [\apj] {10.1086/169789}, \href
  {https://ui.adsabs.harvard.edu/abs/1991ApJ...370....1M} {370, 1}

\bibitem[\protect\citeauthoryear{{Miyazaki} et~al.,}{{Miyazaki}
  et~al.}{2018}]{Miyazaki:2018}
{Miyazaki} S.,  et~al., 2018, \mn@doi [\pasj] {10.1093/pasj/psx063}, \href
  {https://ui.adsabs.harvard.edu/abs/2018PASJ...70S...1M} {70, S1}

\bibitem[\protect\citeauthoryear{{Moles} et~al.,}{{Moles}
  et~al.}{2008}]{Moles:2008}
{Moles} M.,  et~al., 2008, \mn@doi [\aj] {10.1088/0004-6256/136/3/1325}, \href
  {https://ui.adsabs.harvard.edu/abs/2008AJ....136.1325M} {136, 1325}

\bibitem[\protect\citeauthoryear{{Moore}}{{Moore}}{1994}]{Moore:1994}
{Moore} B.,  1994, \mn@doi [\nat] {10.1038/370629a0}, \href
  {https://ui.adsabs.harvard.edu/abs/1994Natur.370..629M} {370, 629}

\bibitem[\protect\citeauthoryear{{Moore}, {Ghigna}, {Governato}, {Lake},
  {Quinn}, {Stadel}  \& {Tozzi}}{{Moore} et~al.}{1999}]{Moore:1999}
{Moore} B.,  {Ghigna} S.,  {Governato} F.,  {Lake} G.,  {Quinn} T.,  {Stadel}
  J.,   {Tozzi} P.,  1999, \mn@doi [\apjl] {10.1086/312287}, \href
  {https://ui.adsabs.harvard.edu/abs/1999ApJ...524L..19M} {524, L19}

\bibitem[\protect\citeauthoryear{{Munshi}, {Brooks}, {Applebaum},
  {Christensen}, {Quinn}  \& {Sligh}}{{Munshi} et~al.}{2021}]{Munshi:2021}
{Munshi} F.,  {Brooks} A.~M.,  {Applebaum} E.,  {Christensen} C.~R.,  {Quinn}
  T.,   {Sligh} S.,  2021, \mn@doi [\apj] {10.3847/1538-4357/ac0db6}, \href
  {https://ui.adsabs.harvard.edu/abs/2021ApJ...923...35M} {923, 35}

\bibitem[\protect\citeauthoryear{{Nadler} et~al.,}{{Nadler}
  et~al.}{2020}]{Nadler:2020}
{Nadler} E.~O.,  et~al., 2020, \mn@doi [\apj] {10.3847/1538-4357/ab846a}, \href
  {https://ui.adsabs.harvard.edu/abs/2020ApJ...893...48N} {893, 48}

\bibitem[\protect\citeauthoryear{{Nadler}, {Banerjee}, {Adhikari}, {Mao}  \&
  {Wechsler}}{{Nadler} et~al.}{2021}]{Nadler:2021}
{Nadler} E.~O.,  {Banerjee} A.,  {Adhikari} S.,  {Mao} Y.-Y.,   {Wechsler}
  R.~H.,  2021, \mn@doi [\apjl] {10.3847/2041-8213/ac29c1}, \href
  {https://ui.adsabs.harvard.edu/abs/2021ApJ...920L..11N} {920, L11}

\bibitem[\protect\citeauthoryear{{Newman} et~al.,}{{Newman}
  et~al.}{2013}]{Newman:2013}
{Newman} J.~A.,  et~al., 2013, \mn@doi [\apjs] {10.1088/0067-0049/208/1/5},
  \href {https://ui.adsabs.harvard.edu/abs/2013ApJS..208....5N} {208, 5}

\bibitem[\protect\citeauthoryear{{Oh} et~al.,}{{Oh} et~al.}{2015}]{Oh:2015}
{Oh} S.-H.,  et~al., 2015, \mn@doi [\aj] {10.1088/0004-6256/149/6/180}, \href
  {https://ui.adsabs.harvard.edu/abs/2015AJ....149..180O} {149, 180}

\bibitem[\protect\citeauthoryear{{Oke} \& {Gunn}}{{Oke} \&
  {Gunn}}{1983}]{Oke1983}
{Oke} J.~B.,  {Gunn} J.~E.,  1983, \mn@doi [\apj] {10.1086/160817}, \href
  {https://ui.adsabs.harvard.edu/abs/1983ApJ...266..713O} {266, 713}

\bibitem[\protect\citeauthoryear{{Oliver} et~al.,}{{Oliver}
  et~al.}{2012}]{Oliver:2012}
{Oliver} S.~J.,  et~al., 2012, \mn@doi [\mnras]
  {10.1111/j.1365-2966.2012.20912.x}, \href
  {https://ui.adsabs.harvard.edu/abs/2012MNRAS.424.1614O} {424, 1614}

\bibitem[\protect\citeauthoryear{{Oman} et~al.,}{{Oman}
  et~al.}{2015}]{Oman:2015}
{Oman} K.~A.,  et~al., 2015, \mn@doi [\mnras] {10.1093/mnras/stv1504}, \href
  {https://ui.adsabs.harvard.edu/abs/2015MNRAS.452.3650O} {452, 3650}

\bibitem[\protect\citeauthoryear{{Oman}, {Navarro}, {Sales}, {Fattahi},
  {Frenk}, {Sawala}, {Schaller}  \& {White}}{{Oman} et~al.}{2016}]{Oman:2016}
{Oman} K.~A.,  {Navarro} J.~F.,  {Sales} L.~V.,  {Fattahi} A.,  {Frenk} C.~S.,
  {Sawala} T.,  {Schaller} M.,   {White} S. D.~M.,  2016, \mn@doi [\mnras]
  {10.1093/mnras/stw1251}, \href
  {https://ui.adsabs.harvard.edu/abs/2016MNRAS.460.3610O} {460, 3610}

\bibitem[\protect\citeauthoryear{{Ouchi} et~al.,}{{Ouchi}
  et~al.}{2018}]{Ouchi:2018}
{Ouchi} M.,  et~al., 2018, \mn@doi [\pasj] {10.1093/pasj/psx074}, \href
  {https://ui.adsabs.harvard.edu/abs/2018PASJ...70S..13O} {70, S13}

\bibitem[\protect\citeauthoryear{{Papastergis}, {Giovanelli}, {Haynes}  \&
  {Shankar}}{{Papastergis} et~al.}{2015}]{Papastergis:2015}
{Papastergis} E.,  {Giovanelli} R.,  {Haynes} M.~P.,   {Shankar} F.,  2015,
  \mn@doi [\aap] {10.1051/0004-6361/201424909}, \href
  {https://ui.adsabs.harvard.edu/abs/2015A&A...574A.113P} {574, A113}

\bibitem[\protect\citeauthoryear{{Pontzen} \& {Governato}}{{Pontzen} \&
  {Governato}}{2012}]{Pontzen:2012}
{Pontzen} A.,  {Governato} F.,  2012, \mn@doi [\mnras]
  {10.1111/j.1365-2966.2012.20571.x}, \href
  {https://ui.adsabs.harvard.edu/abs/2012MNRAS.421.3464P} {421, 3464}

\bibitem[\protect\citeauthoryear{{Posti}, {Marasco}, {Fraternali}  \&
  {Famaey}}{{Posti} et~al.}{2019}]{Posti:2019}
{Posti} L.,  {Marasco} A.,  {Fraternali} F.,   {Famaey} B.,  2019, \mn@doi
  [\aap] {10.1051/0004-6361/201935982}, \href
  {https://ui.adsabs.harvard.edu/abs/2019A&A...629A..59P} {629, A59}

\bibitem[\protect\citeauthoryear{{Read} \& {Erkal}}{{Read} \&
  {Erkal}}{2019}]{Read:2019b}
{Read} J.~I.,  {Erkal} D.,  2019, \mn@doi [\mnras] {10.1093/mnras/stz1320},
  \href {https://ui.adsabs.harvard.edu/abs/2019MNRAS.487.5799R} {487, 5799}

\bibitem[\protect\citeauthoryear{{Read}, {Wilkinson}, {Evans}, {Gilmore}  \&
  {Kleyna}}{{Read} et~al.}{2006}]{Read:2006}
{Read} J.~I.,  {Wilkinson} M.~I.,  {Evans} N.~W.,  {Gilmore} G.,   {Kleyna}
  J.~T.,  2006, \mn@doi [\mnras] {10.1111/j.1365-2966.2005.09959.x}, \href
  {https://ui.adsabs.harvard.edu/abs/2006MNRAS.367..387R} {367, 387}

\bibitem[\protect\citeauthoryear{{Read}, {Iorio}, {Agertz}  \&
  {Fraternali}}{{Read} et~al.}{2016}]{Read:2016}
{Read} J.~I.,  {Iorio} G.,  {Agertz} O.,   {Fraternali} F.,  2016, \mn@doi
  [\mnras] {10.1093/mnras/stw1876}, \href
  {https://ui.adsabs.harvard.edu/abs/2016MNRAS.462.3628R} {462, 3628}

\bibitem[\protect\citeauthoryear{{Read}, {Iorio}, {Agertz}  \&
  {Fraternali}}{{Read} et~al.}{2017}]{Read:2017}
{Read} J.~I.,  {Iorio} G.,  {Agertz} O.,   {Fraternali} F.,  2017, \mn@doi
  [\mnras] {10.1093/mnras/stx147}, \href
  {https://ui.adsabs.harvard.edu/abs/2017MNRAS.467.2019R} {467, 2019}

\bibitem[\protect\citeauthoryear{{Read}, {Walker}  \& {Steger}}{{Read}
  et~al.}{2019}]{Read:2019a}
{Read} J.~I.,  {Walker} M.~G.,   {Steger} P.,  2019, \mn@doi [\mnras]
  {10.1093/mnras/sty3404}, \href
  {https://ui.adsabs.harvard.edu/abs/2019MNRAS.484.1401R} {484, 1401}

\bibitem[\protect\citeauthoryear{{Relatores} et~al.,}{{Relatores}
  et~al.}{2019}]{Relatores:2019}
{Relatores} N.~C.,  et~al., 2019, \mn@doi [\apj] {10.3847/1538-4357/ab5305},
  \href {https://ui.adsabs.harvard.edu/abs/2019ApJ...887...94R} {887, 94}

\bibitem[\protect\citeauthoryear{{Robles} et~al.,}{{Robles}
  et~al.}{2017}]{Robles:2017}
{Robles} V.~H.,  et~al., 2017, \mn@doi [\mnras] {10.1093/mnras/stx2253}, \href
  {https://ui.adsabs.harvard.edu/abs/2017MNRAS.472.2945R} {472, 2945}

\bibitem[\protect\citeauthoryear{{Roper}, {Oman}, {Frenk},
  {Ben{\'\i}tez-Llambay}, {Navarro}  \& {Santos-Santos}}{{Roper}
  et~al.}{2023}]{Roper:2023}
{Roper} F.~A.,  {Oman} K.~A.,  {Frenk} C.~S.,  {Ben{\'\i}tez-Llambay} A.,
  {Navarro} J.~F.,   {Santos-Santos} I. M.~E.,  2023, \mn@doi [\mnras]
  {10.1093/mnras/stad549}, \href
  {https://ui.adsabs.harvard.edu/abs/2023MNRAS.521.1316R} {521, 1316}

\bibitem[\protect\citeauthoryear{{Rowe} et~al.,}{{Rowe}
  et~al.}{2015}]{Rowe:2015}
{Rowe} B.~T.~P.,  et~al., 2015, \mn@doi [Astronomy and Computing]
  {10.1016/j.ascom.2015.02.002}, \href
  {https://ui.adsabs.harvard.edu/abs/2015A&C....10..121R} {10, 121}

\bibitem[\protect\citeauthoryear{{Rubin}, {Ford}  \& {Thonnard}}{{Rubin}
  et~al.}{1978}]{Rubin:1978}
{Rubin} V.~C.,  {Ford} W.~K. J.,   {Thonnard} N.,  1978, \mn@doi [\apjl]
  {10.1086/182804}, \href
  {https://ui.adsabs.harvard.edu/abs/1978ApJ...225L.107R} {225, L107}

\bibitem[\protect\citeauthoryear{{Rubin}, {Ford}  \& {Thonnard}}{{Rubin}
  et~al.}{1980}]{Rubin:1980}
{Rubin} V.~C.,  {Ford} W.~K. J.,   {Thonnard} N.,  1980, \mn@doi [\apj]
  {10.1086/158003}, \href
  {https://ui.adsabs.harvard.edu/abs/1980ApJ...238..471R} {238, 471}

\bibitem[\protect\citeauthoryear{{Sales}, {Wetzel}  \& {Fattahi}}{{Sales}
  et~al.}{2022}]{Sales:2022}
{Sales} L.~V.,  {Wetzel} A.,   {Fattahi} A.,  2022, \mn@doi [Nature Astronomy]
  {10.1038/s41550-022-01689-w}, \href
  {https://ui.adsabs.harvard.edu/abs/2022NatAs...6..897S} {6, 897}

\bibitem[\protect\citeauthoryear{{S{\'a}nchez-Bl{\'a}zquez}
  et~al.,}{{S{\'a}nchez-Bl{\'a}zquez} et~al.}{2006}]{Sanchez-Blazquez:2006}
{S{\'a}nchez-Bl{\'a}zquez} P.,  et~al., 2006, \mn@doi [\mnras]
  {10.1111/j.1365-2966.2006.10699.x}, \href
  {https://ui.adsabs.harvard.edu/abs/2006MNRAS.371..703S} {371, 703}

\bibitem[\protect\citeauthoryear{{Santos-Santos}, {Di Cintio}, {Brook},
  {Macci{\`o}}, {Dutton}  \& {Dom{\'\i}nguez-Tenreiro}}{{Santos-Santos}
  et~al.}{2018}]{Santos-Santos:2018}
{Santos-Santos} I.~M.,  {Di Cintio} A.,  {Brook} C.~B.,  {Macci{\`o}} A.,
  {Dutton} A.,   {Dom{\'\i}nguez-Tenreiro} R.,  2018, \mn@doi [\mnras]
  {10.1093/mnras/stx2660}, \href
  {https://ui.adsabs.harvard.edu/abs/2018MNRAS.473.4392S} {473, 4392}

\bibitem[\protect\citeauthoryear{{Scoville} et~al.,}{{Scoville}
  et~al.}{2007}]{Scoville:2007}
{Scoville} N.,  et~al., 2007, \mn@doi [\apjs] {10.1086/516585}, \href
  {https://ui.adsabs.harvard.edu/abs/2007ApJS..172....1S} {172, 1}

\bibitem[\protect\citeauthoryear{{Singh}, {Mandelbaum}, {Seljak}, {Slosar}  \&
  {Vazquez Gonzalez}}{{Singh} et~al.}{2017}]{Singh:2017}
{Singh} S.,  {Mandelbaum} R.,  {Seljak} U.,  {Slosar} A.,   {Vazquez Gonzalez}
  J.,  2017, \mn@doi [\mnras] {10.1093/mnras/stx1828}, \href
  {https://ui.adsabs.harvard.edu/abs/2017MNRAS.471.3827S} {471, 3827}

\bibitem[\protect\citeauthoryear{{Smith} et~al.,}{{Smith}
  et~al.}{2003}]{Smith:2003}
{Smith} R.~E.,  et~al., 2003, \mn@doi [\mnras]
  {10.1046/j.1365-8711.2003.06503.x}, \href
  {https://ui.adsabs.harvard.edu/abs/2003MNRAS.341.1311S} {341, 1311}

\bibitem[\protect\citeauthoryear{{Speagle}}{{Speagle}}{2020}]{Speagle:2020}
{Speagle} J.~S.,  2020, \mn@doi [\mnras] {10.1093/mnras/staa278}, \href
  {https://ui.adsabs.harvard.edu/abs/2020MNRAS.493.3132S} {493, 3132}

\bibitem[\protect\citeauthoryear{{Speagle} et~al.,}{{Speagle}
  et~al.}{2019}]{Speagle:2019}
{Speagle} J.~S.,  et~al., 2019, \mn@doi [\mnras] {10.1093/mnras/stz2968}, \href
  {https://ui.adsabs.harvard.edu/abs/2019MNRAS.490.5658S} {490, 5658}

\bibitem[\protect\citeauthoryear{{Spergel} et~al.,}{{Spergel}
  et~al.}{2013}]{Spergel:2013}
{Spergel} D.,  et~al., 2013, arXiv e-prints, \href
  {https://ui.adsabs.harvard.edu/abs/2013arXiv1305.5425S} {p. arXiv:1305.5425}

\bibitem[\protect\citeauthoryear{{Takahashi}, {Sato}, {Nishimichi}, {Taruya}
  \& {Oguri}}{{Takahashi} et~al.}{2012}]{Takahashi:2012}
{Takahashi} R.,  {Sato} M.,  {Nishimichi} T.,  {Taruya} A.,   {Oguri} M.,
  2012, \mn@doi [\apj] {10.1088/0004-637X/761/2/152}, \href
  {https://ui.adsabs.harvard.edu/abs/2012ApJ...761..152T} {761, 152}

\bibitem[\protect\citeauthoryear{{Tanaka} et~al.,}{{Tanaka}
  et~al.}{2018}]{Tanaka:2018}
{Tanaka} M.,  et~al., 2018, \mn@doi [\pasj] {10.1093/pasj/psx077}, \href
  {https://ui.adsabs.harvard.edu/abs/2018PASJ...70S...9T} {70, S9}

\bibitem[\protect\citeauthoryear{{Taniguchi} et~al.,}{{Taniguchi}
  et~al.}{2007}]{Taniguchi:2007}
{Taniguchi} Y.,  et~al., 2007, \mn@doi [\apjs] {10.1086/516596}, \href
  {https://ui.adsabs.harvard.edu/abs/2007ApJS..172....9T} {172, 9}

\bibitem[\protect\citeauthoryear{{Taniguchi} et~al.,}{{Taniguchi}
  et~al.}{2015}]{Taniguchi:2015}
{Taniguchi} Y.,  et~al., 2015, \mn@doi [\pasj] {10.1093/pasj/psv106}, \href
  {https://ui.adsabs.harvard.edu/abs/2015PASJ...67..104T} {67, 104}

\bibitem[\protect\citeauthoryear{{Taylor}}{{Taylor}}{2005}]{topcat}
{Taylor} M.~B.,  2005, in {Shopbell} P.,  {Britton} M.,   {Ebert} R.,  eds,
  Astronomical Society of the Pacific Conference Series Vol. 347, Astronomical
  Data Analysis Software and Systems XIV. p.~29

\bibitem[\protect\citeauthoryear{{Teyssier}, {Pontzen}, {Dubois}  \&
  {Read}}{{Teyssier} et~al.}{2013}]{Teyssier:2013}
{Teyssier} R.,  {Pontzen} A.,  {Dubois} Y.,   {Read} J.~I.,  2013, \mn@doi
  [\mnras] {10.1093/mnras/sts563}, \href
  {https://ui.adsabs.harvard.edu/abs/2013MNRAS.429.3068T} {429, 3068}

\bibitem[\protect\citeauthoryear{{Tollet} et~al.,}{{Tollet}
  et~al.}{2016}]{Tollet:2016}
{Tollet} E.,  et~al., 2016, \mn@doi [\mnras] {10.1093/mnras/stv2856}, \href
  {https://ui.adsabs.harvard.edu/abs/2016MNRAS.456.3542T} {456, 3542}

\bibitem[\protect\citeauthoryear{Virtanen et~al.,}{Virtanen
  et~al.}{2020}]{SciPy}
Virtanen P.,  et~al., 2020, \mn@doi [Nature Methods]
  {10.1038/s41592-019-0686-2}, \href {https://rdcu.be/b08Wh} {17, 261}

\bibitem[\protect\citeauthoryear{{Wetzel}, {Hopkins}, {Kim},
  {Faucher-Gigu{\`e}re}, {Kere{\v{s}}}  \& {Quataert}}{{Wetzel}
  et~al.}{2016}]{Wetzel:2016}
{Wetzel} A.~R.,  {Hopkins} P.~F.,  {Kim} J.-h.,  {Faucher-Gigu{\`e}re} C.-A.,
  {Kere{\v{s}}} D.,   {Quataert} E.,  2016, \mn@doi [\apjl]
  {10.3847/2041-8205/827/2/L23}, \href
  {https://ui.adsabs.harvard.edu/abs/2016ApJ...827L..23W} {827, L23}

\bibitem[\protect\citeauthoryear{{Wilson}, {Kaiser}, {Luppino}  \&
  {Cowie}}{{Wilson} et~al.}{2001}]{Wilson:2001}
{Wilson} G.,  {Kaiser} N.,  {Luppino} G.~A.,   {Cowie} L.~L.,  2001, \mn@doi
  [\apj] {10.1086/321441}, \href
  {https://ui.adsabs.harvard.edu/abs/2001ApJ...555..572W} {555, 572}

\bibitem[\protect\citeauthoryear{{York} et~al.,}{{York}
  et~al.}{2000}]{York:2000}
{York} D.~G.,  et~al., 2000, \mn@doi [\aj] {10.1086/301513}, \href
  {https://ui.adsabs.harvard.edu/abs/2000AJ....120.1579Y} {120, 1579}

\bibitem[\protect\citeauthoryear{{Zamojski} et~al.,}{{Zamojski}
  et~al.}{2007}]{Zamojski:2007}
{Zamojski} M.~A.,  et~al., 2007, \mn@doi [\apjs] {10.1086/516593}, \href
  {https://ui.adsabs.harvard.edu/abs/2007ApJS..172..468Z} {172, 468}

\bibitem[\protect\citeauthoryear{{Zheng} et~al.,}{{Zheng}
  et~al.}{2017}]{Zheng:2017}
{Zheng} Z.-Y.,  et~al., 2017, \mn@doi [\apjl] {10.3847/2041-8213/aa794f}, \href
  {https://ui.adsabs.harvard.edu/abs/2017ApJ...842L..22Z} {842, L22}

\bibitem[\protect\citeauthoryear{{Zheng} et~al.,}{{Zheng}
  et~al.}{2019}]{Zheng:2019}
{Zheng} Z.-Y.,  et~al., 2019, \mn@doi [\pasp] {10.1088/1538-3873/ab1c32}, \href
  {https://ui.adsabs.harvard.edu/abs/2019PASP..131g4502Z} {131, 074502}

\makeatother
\end{thebibliography}

\appendix
\label{appendix}

\label{lastpage}


\end{document}